\documentclass[conference,10pt]{IEEEtran}
\pdfoutput=1

\usepackage{algorithm, balance}
\usepackage{algorithmicx}
\usepackage[noend]{algpseudocode}
\usepackage[pdftex]{graphicx}
\usepackage{epstopdf}
\usepackage{xcolor}
\usepackage{xspace}
\usepackage[final]{pdfpages}
\usepackage{times}
\def\iscomments{0}
\def\isarxiv{1}

\newif\ifanonymous
\newif\ifcomments
\newif\iftechnicalreport
\newif\ifarxiv
\newif\ifsecon

\ifdefined\isanonymous
\anonymoustrue
\fi

\ifdefined\iscomments
\else
\commentstrue
\fi

\ifdefined\isarxiv
\arxivtrue
\else
\arxivfalse
\fi

\ifdefined\istechnicalreport
\technicalreporttrue
\else
\technicalreportfalse
\fi

\ifdefined\issecon
\secontrue
\fi

\ifcomments
\usepackage{changes}
\else
\usepackage[final]{changes}
\fi

\definechangesauthor[color=green]{CV}
\definechangesauthor[color=blue]{MJ}
\definechangesauthor[color=red]{NEW}

\usepackage[cmex10]{amsmath}

\usepackage{import}

\makeindex
\begin{document}

\newcommand{\sysname}{\texttt{STASH}\xspace}
\newcommand{\myparagraph}[1]{\noindent\textbf{#1.}}

\newcommand{\ver}{$\mathcal{V}$\xspace}
\newcommand{\pro}{$\mathcal{P}$\xspace}
\newcommand{\user}{$\mathcal{U}$\xspace}
\newcommand{\adv}{$\mathcal{A}$\xspace}
\newcommand*\Let[2]{\State #1 $\gets$ #2}

\title{\sysname: Securing transparent authentication schemes using prover-side
proximity verification}

\ifanonymous
\author{
}
\else
\author{\IEEEauthorblockN{Mika Juuti\IEEEauthorrefmark{1},
Christian Vaas\IEEEauthorrefmark{2},
Ivo Sluganovic\IEEEauthorrefmark{2},
Hans Liljestrand\IEEEauthorrefmark{3},
N. Asokan\IEEEauthorrefmark{3} and
Ivan Martinovic\IEEEauthorrefmark{2}}
\IEEEauthorblockA{\IEEEauthorrefmark{1}Aalto University, mika.juuti@aalto.fi}
\IEEEauthorblockA{\IEEEauthorrefmark{2}University of Oxford, christian.vaas@cs.ox.ac.uk, ivo.sluganovic@cs.ox.ac.uk, ivan.martinovic@cs.ox.ac.uk}
\IEEEauthorblockA{\IEEEauthorrefmark{3}Aalto University and University
  of Helsinki, hans.liljestrand@aalto.fi, asokan@acm.org}}
\fi

\maketitle

\begin{abstract}

Transparent authentication (TA) schemes are those in which a user is authenticated by a verifier \emph{without requiring explicit user interaction}. By doing so, those schemes promise high usability and security simultaneously. The majority of TA implementations rely on the received signal strength as an indicator for the proximity of a user device (prover). However, such implicit proximity verification is not secure against an adversary who can relay messages over a larger distance.

\noindent
In this paper, we propose a novel approach for thwarting relay attacks in TA schemes: the prover permits access to authentication credentials only if it can confirm that it is near the verifier. We present \sysname, a system for relay-resilient transparent authentication in which the prover does proximity verification by comparing its approach trajectory towards the intended verifier with known authorized reference trajectories. Trajectories are measured using low-cost sensors commonly available on personal devices. We demonstrate the security of \sysname against a class of adversaries and its ease-of-use by analyzing empirical data, collected using a \sysname prototype. \sysname is efficient and can be easily integrated to complement existing TA schemes.

\end{abstract}

\section{Introduction}

User authentication is necessary in order to regulate access to data or physical objects. The predominant approach still relies on the use of passwords which suffers from drawbacks in terms of both usability and security~\cite{DBLP:journals/cacm/BonneauHOS15}. Effective and secure alternatives to password-based authentication have yet to emerge~\cite{DBLP:conf/sp/BonneauHOS12}. This has sparked the development of \emph{transparent authentication} (TA) with systems exploiting characteristic cues such as behavior~\cite{Xiao2016}, biometrics~\cite{Tefas2001}, or environmental context~\cite{Truong2014}.
Zero-interaction authentication (ZIA) is a class of TA schemes~\cite{Corner2002} that relies on a \emph{verifier} \ver to authenticate a user when a \emph{prover} device \pro associated with the user is nearby. In ZIA schemes \ver typically verifies the proximity of \pro by measuring either the strength of radio signals emitted from \pro over some short range wireless channel, e.g. BlueProximity~\cite{Project}, or the time required to transmit messages over that channel, e.g. ``keyless entry and start'' systems for cars. However, these TA schemes remain vulnerable to \emph{relay attacks}~\cite{Francillon2011,Francis2012},
where the attacker relays messages between \pro and \ver when they are not co-located, leading to \ver falsely concluding that \pro is nearby.

There are several known defense techniques against these attacks such as distance bounding
protocols~\cite{Brands1993} and comparison of ambient contexts between \pro and \ver~\cite{Halevi2012,Truong2014}. However, these methods are faced with deployment challenges. In effect, context-based relay attack defense systems have unclear security
guarantees~\cite{Shrestha151104}. Additionally, distance bounding methods require precise timing, and
1 ns measurement error can affect the estimated distance by 30
cm. Therefore, additional hardware and low-level software changes seem necessary.

In this paper we propose a novel approach to thwart relay attacks on proximity-based transparent authentication systems. We present \sysname, a system that enforces proximity verification by \pro to an intended \ver before \emph{allowing access} to the credentials used in the authentication protocol. The method uses \pro's on-board micro-electromechanical system (MEMS) sensors to measure its \emph{approach trajectory} towards \ver and compares it with authorized reference paths.
A central design principle in \sysname is to rely only on sensors monitoring \pro's own movement (e.g. accelerometer and gyroscope) rather than on sensors that measure environmental factors that can be manipulated or falsified (e.g. GPS, radio signal emission, or ambient properties). We built \sysname as an Android application and used it to gather trajectory data of 20 different routes in two cities (totaling 123 km). Using this dataset, we demonstrate that \sysname has acceptable false reject (FRR) and false accept (FAR) rates.

Commodity devices provide low-cost MEMS sensors that are noisy, include bias terms and miss data. Designing \sysname to work on commodity devices raised several technical challenges leading to questions like ``how to effectively represent a trajectory using only accelerometer/gyroscope measurements?'' and ``how to best compare two trajectories?''. Additionally, an energy budget examination of portable devices is necessary to understand the feasibility of \sysname. In this paper, we address these challenges and evaluate the resulting system systemically. Briefly, our contributions are the following:

\begin{itemize}
\itemsep0em
  \item We propose using \textbf{prover-side proximity verification to resist relay attacks} against proximity-based transparent authentication systems (Sections~\ref{sec:conceptAndAssumptions} and \ref{sec:implementation}).
  \item We design and implement a \textbf{concrete system, \sysname}, incorporating this idea by addressing several challenges in measuring prover's approach trajectory and using it to determine proximity to verifier (Section~\ref{sec:implementation}).
  \item By systematically analyzing trajectory data in two cities, we \textbf{demonstrate the security and usability of \sysname} (Section~\ref{sec:experimentalEvaluation}). We also show that \sysname's average \textbf{energy consumption is low}: we estimate that under typical usage conditions, the battery drain due to \sysname over the course of a work day is in the range of 4\%-7\% of battery capacity.
(Section~\ref{sec:battery}).
\end{itemize}

\section{Concept and Assumptions}\label{sec:conceptAndAssumptions}

\subsection{System Model}\label{sec:systemModel}

\begin{figure}[htbp]
    \centering
    \def\svgwidth{0.85\columnwidth}
    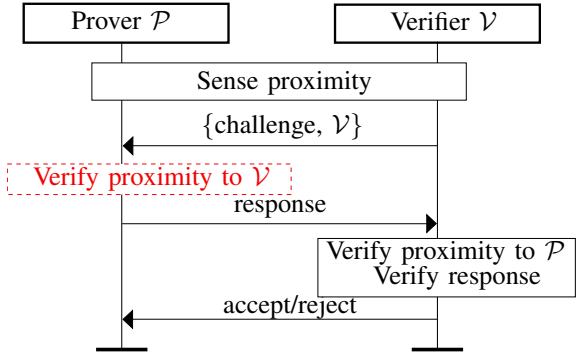
		\caption{Transparent authentication: challenge-response protocol triggered upon sensing proximity of \pro to \ver. The proximity verification component introduced in our approach is shown as a red dashed box.}
		\label{fig:system_model}
\end{figure}

Figure~\ref{fig:system_model} illustrates the basis of proximity-based transparent authentication. The goal of this model is to enable confirmation from the verifier \ver that a user \user is nearby. For this purpose, \user has a personal device \pro and authentication is then based on a challenge-response protocol using a previously established security association, e.g. a shared symmetric key between \pro and \ver.
However, in addition to verifying authenticity of \pro, \ver also verifies its proximity to \pro, a process which is vulnerable to relay attacks.

To protect against those attacks, we improve this model by having \pro regulate access to the authentication credentials through first verifying its proximity to \ver. In particular, we propose that \pro does this proximity verification by examining its approach trajectory towards \ver. If proximity verification fails, \user is asked to explicitly confirm she is near \ver.

\myparagraph{Application scenario: premise access control} We target a scenario where a person \user accompanied by \pro routinely approaches an access controlling barrier \ver. After transparently authenticating \pro, \ver will open the barrier to let \user pass. If proximity verification fails, \user may still open the gate through explicit proximity confirmation using \pro. Examples for such scenario are: a person \user on a vehicle \pro such as a bicycle, wheelchair or car requiring easy and fast access through a gate or door \ver at her home or workplace.

\subsection{Adversary Model}\label{sec:threatModel}
We consider an adversary \adv who has deployed a wireless relay providing him with Dolev-Yao~\cite{dolev1983security} capabilities. Although \adv can control the message flow between \pro and \ver, he cannot break the cryptographic protection of a secured channel. Nevertheless, by relaying the challenge and response and thus artificially extending the range of the wireless channel, \adv can successfully bypass the proximity verification since \ver will be measuring the relay device's signal strength rather than \pro's. Even if time of flight is used to estimate the distance, off-the-shelf hardware is not precise enough to provide a secure estimate of the distance between two devices. Attacks like this are actively being exploited as demonstrated for several scenarios~\cite{Francillon2011,Francis2012}.

We do not consider adversaries who gain physical access to \pro. Continuous user authentication techniques, such as biometric authentication~\cite{Patel2016} can ensure it is the legitimate user who is in possession of the the device. We also assume that the device has not been infected with malware which can be ensured using platform security and anti-malware tools.

\subsection{Design Goals and Challenges}\label{sec:designGoals}

\myparagraph{Goals}
We set the following goals for our relay-resilient proximity verification system:
\begin{enumerate}
  \renewcommand{\labelenumi}{R\arabic{enumi}. }
\item \label{R-usability}\textbf{Usability}: Transparent authentication must minimize explicit user action. If trajectory comparison fails when \user is in fact near \ver, \user will be required to fall back to explicit proximity confirmation. Our system should therefore minimize the false reject rate.
\item \label{R-security}\textbf{Security}: The system should not incorrectly conclude that \user is near \ver even in the presence of a relay. Therefore it should minimize the false accept rate.
\item \label{R-efficiency}\textbf{Efficiency}: The computational and energy costs of proximity verification should be small to not diminish the user experience.
\item \label{R-no-ext-sig}\textbf{No external signals}: Since \adv can control ambient properties, proximity verification should not depend on any external signals.
\item \label{R-local-decisions}\textbf{Local decision-making}: Proximity verification must be carried out entirely within \pro.
\end{enumerate}

There are two rationales for R\ref{R-local-decisions}. One is \textbf{privacy}: data collected for proximity verification should not be exposed to any third party. The other is \textbf{deployability}: a local solution can be seamlessly integrated into any proximity-based transparent authentication scheme by only modifying \pro without having to change the protocol and thus the implementation of \ver.

\section{\sysname Architecture}\label{sec:implementation}

We now describe \sysname, our system that uses prover-side proximity verification to prevent relay attacks.

\subsection{Trajectory Representation}
\label{subsec:trajectory}

\begin{figure}[htbp]
\centering
\includegraphics[width=.5\columnwidth]{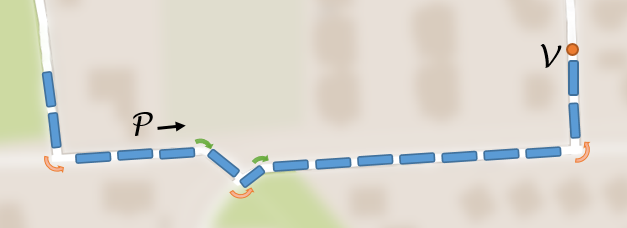}
\caption{\pro's approach trajectory towards \ver is described by a set of primitives needed to reach \ver. In this example, the primitives are \textit{move 2, left 90$^{\circ}$, move 3, right 30$^{\circ}$, move 1, left 90$^{\circ}$, move 1, right 30$^{\circ}$, move 7, left 90$^{\circ}$, move 2.}}
\label{fig:schematic}
\end{figure}

To satisfy requirement~R\ref{R-no-ext-sig}, we avoid external, insecure, data sources like GPS~\cite{Tippenhauer2011} or ambient sensor modalities and rely only on gyroscope and accelerometer to capture user's movement. We represent a trajectory as a \emph{temporally ordered sequence} of discrete primitives consisting of segments of movement interleaved with left or right turns derived from angular information.
(See Figure~\ref{fig:schematic} for an example.)

An intuitive way to represent a trajectory is as a sequence of coordinates, like in dead reckoning~\cite{jimenez2009comparison}.
However, a one-dimensional sequence of primitives is more robust to sensor noise than a two-(or even three-) dimensional coordinate: the impact of a missed turn on the resulting sequence is less than it is on the result of a dead reckoning algorithm.

Using sensor data, we recognize two streams of primitives: 
$(M/S,t_i)$ symbols
(for ``movement'' or ``stationary'' at time $t_i$)
are generated at a fixed rate and $(L/R,t_i)$ symbols
(turn ``left'' or ``right'' at time $t_i$)
are generated opportunistically whenever a turn is detected. The two streams are then combined into one sequence, with turn events taking precedence. The overall system has three essential parts: primitive generation, trajectory comparison and authorized trajectory updating.

\subsection{Primitive Generation}
\label{sec:event_generation}

\myparagraph{Turn primitives} By exploring different sampling rates and sensors, we concluded that a 20 Hz sampling rate is sufficient to detect turns with a precision of 15$^{\circ}$.
To achieve this, we project the gyroscope to ground direction, obtain the heading angle by integrating the angular speed and then record turns when the 2s sliding window standard deviation of the heading angle is above a threshold ($\sigma_1 = 3^{\circ}$).
\ifsecon{Orientation changes}\else{However, sudden gravity shifts}\fi\footnote{Gravity in Android is a software sensor (a low-pass filter on raw accelerometer data) which takes a few seconds after an orientation change to stabilize.} 
cause errors in turn estimation: we disregard gyroscope data as unreliable in such situations.
To remove drift in MEMS gyroscopes, we use a high-pass filter, where gyroscope measurements smaller than 
$8.6^{\circ}$/s are exponentially weighted down. Fine-grained beginnings and ends of turns are found where the sliding window standard deviation is above a smaller threshold ($\sigma_2 = 1^{\circ}$).
The turn detection system assumes that \pro has reliable gravity estimates: \pro could for instance be integrated into a vehicle or firmly attached to the body to avoid disturbing the gravity direction. In this paper we collect data by integrating \pro with a bicycle.  

\myparagraph{Movement primitives}
To identify movement, we use a logistic regression (LR) algorithm~\cite{murphy2012machine} that continuously predicts movement mode at one second intervals.
The prediction is done on-the-spot, and does not take previous prediction results into account. In reality however, two successive events are dependent.
We additionally use a Hidden Markov Model (HMM)~\cite{durbin1998biological} to capture this dependency.

In HMMs, probabilities to move between hidden states are modeled with a first-order Markov chain. Each hidden state gives a cue about itself by emitting a primitive at any given time. In our case, we want to determine movement primitives ($M$ or $S$) by observing the output of LR.
We use HMM Viterbi algorithm
\cite{durbin1998biological}
to smoothen the observations into the most likely sequence of primitives. Finally, we determine the representative primitive for each five measurements as the most frequent among the five. 
This scheme gives us a continuous stream of one $M$ or $S$ primitive in five seconds intervals.

\subsection{Trajectory Comparison}
\label{sec:paths}

\begin{figure}[htbp]
    \centering
		\includegraphics[width=0.9\columnwidth]{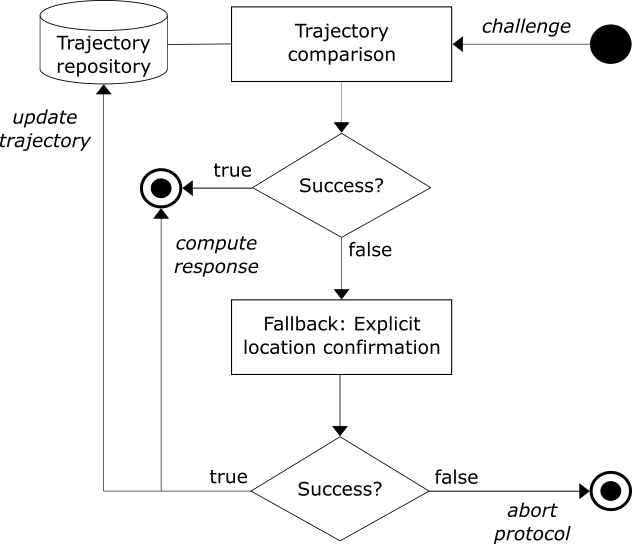}
		\caption{Prover-side proximity verification. \pro ~compares the current candidate path against reference paths in the repository. \pro computes the response to \ver's challenge if the candidate path matches a reference path. The user is asked for explicit proximity confirmation if this check fails.}
	 	\label{fig:access_control}
\end{figure}

We differentiate between two types of trajectories. \emph{Reference paths} are previously authorized trajectories of a \pro towards some verifier \ver . \emph{Candidate paths} are trajectories towards \ver ~perceived by \pro before a proximity verification session. These can contain errors introduced through noisy measurements.
To verify proximity to \ver ,~\pro ~compares the candidate path to the reference path.
Figure~\ref{fig:access_control} shows our proposed proximity verification scheme.
If the trajectory comparison succeeds, \pro continues with the authentication protocol by computing the response to the challenge
and sending it to \ver. If trajectory comparison fails, the system falls back to explicit proximity confirmation by \user. Successful explicit confirmation implies that \pro's
candidate path can be added to the trajectory repository as an authorized trajectory: a reference path.

Candidate and reference paths are represented as sequences of characters as discussed in Section~\ref{subsec:trajectory}. Trajectory comparison is therefore a similarity comparison between strings. We evaluated several string matching metrics and chose to use Needleman-Wunsch (NW) similarity\footnote{Needleman-Wunsch had the best FAR/FRR trade-offs among the tested algorithms on our dataset in Section~\ref{sec:experimentalEvaluation}.}~\cite{durbin1998biological}, which is a combination of the longest common sub-sequence and edit distance algorithms. In NW terminology, both insertions and deletions are called gaps. 
We chose the parametrization: match \textit{+1}, mismatch \textit{-2} and gap \textit{-1}.
Matches can be seen as evidence and mismatches counter-evidence that two sequences are related. Before comparison, $S$ symbols are removed and timestamps ($t_i$) are used to trim strings to the same temporal length.

Instances of the same reference path will differ due to noise from various sources. Therefore, we need to establish a \emph{decision threshold} that determines how much noise is acceptable. If the similarity score is higher than the threshold, the candidate path is accepted. Otherwise it is rejected. This introduces a trade-off between usability (FRR) and security (FAR). An \emph{initial threshold} that has a good FAR/FRR trade-off is determined prior to deployment as discussed in Section~\ref{subsec:thresholds}.

\subsection{Updating Reference Paths}
\label{sec:updates}

Once the system is deployed, we use feedback from failed and successful trajectory comparison attempts to adjust the decision threshold. The initial threshold might under- or overestimate the variation in future instances of a given reference path $r$. The decision threshold should be adjusted to achieve a better FAR/FRR trade-off, either by decreasing (better usability) or increasing (better security) it.

We call such a path-specific decision threshold a \emph{local threshold}. To compute local thresholds for a given $r$ we need instances $i_r$ of $r$ and instances $i_{\bar{r}}$ of reference paths towards other verifiers $\bar{\mathcal{V}}$\xspace.
These are used to calculate \emph{within-} and \emph{between-class similarities}.
When a user \user ~starts using \sysname we do not have enough instances $i_r$, $i_{\bar{r}}$ to compute within- and between-class similarities. Instances $i_r$ will be gradually collected as the user repeatedly traverses $r$. We consider two ways to acquire instances $i_{\bar{r}}$ of paths towards another $\bar{\mathcal{V}}$\xspace:

\begin{itemize}
\itemsep0em
\item Use trajectories generated from a map for the current geographic region.
\item Collect all trajectories of a given user and create a generative probabilistic model (Markov chain) to simulate new reference paths instances.
\end{itemize}

\begin{figure}[tbh]
    \centering
\ifsecon{
    \includegraphics[width=\columnwidth]{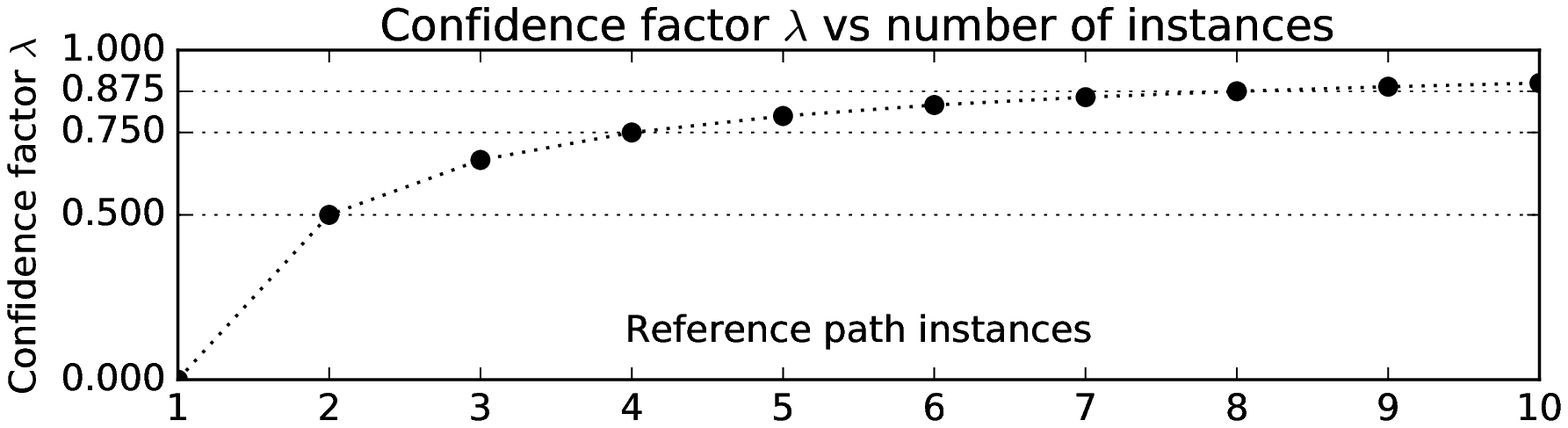}
}\else{
    \includegraphics[width=\columnwidth]{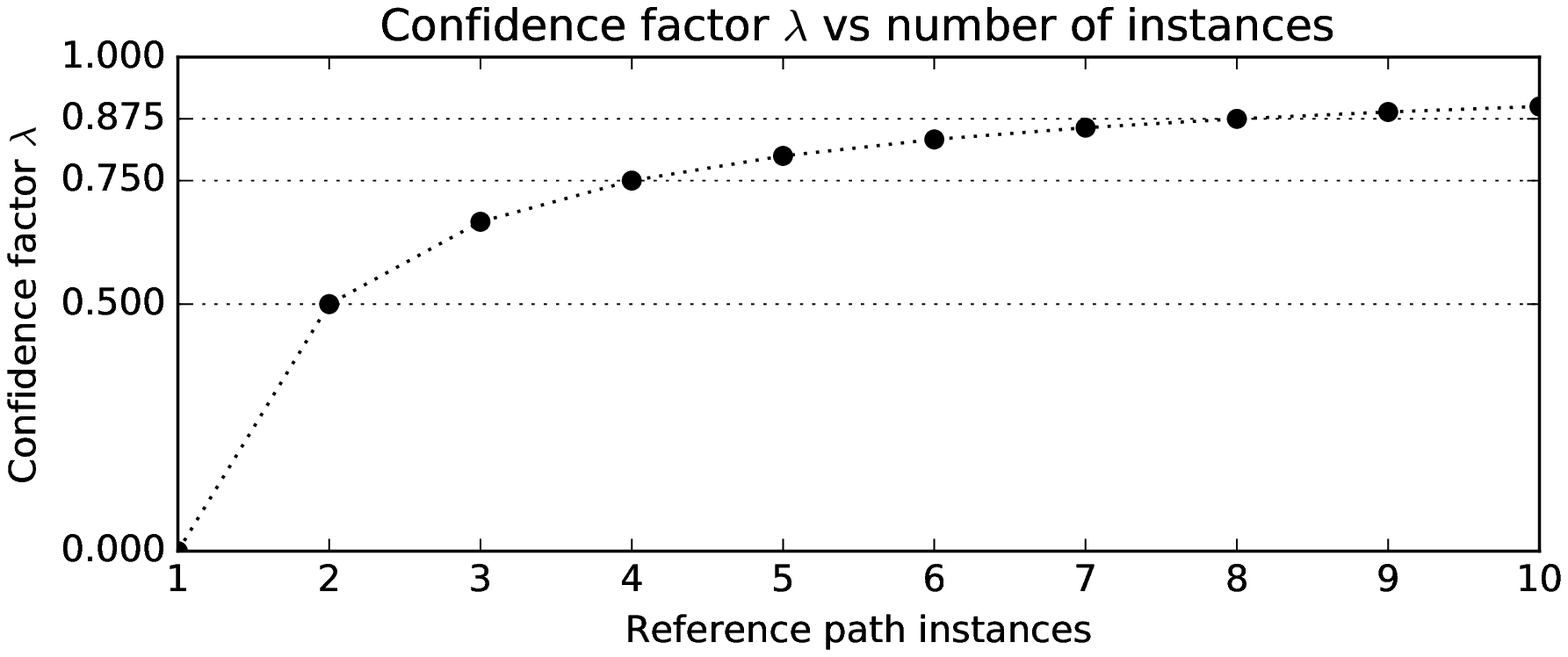}
}\fi
    \caption{The confidence factor $\lambda$ increases with the number of observed reference path instances. The distance to one halves at every multiple of two.}
    \label{fig:cf}
\end{figure}

We choose the latter option in \sysname.
In the beginning, \sysname will only observe a few instances of a reference path and any new local decision threshold we obtain would be severely over-learned.
Naively trusting the seen instances risks a small sample size fallacy. Therefore, we need a way to model the trustworthiness of the estimated local threshold.
We model the confidence as a mixture model using a convex combination~\cite{murphy2012machine} of the thresholds $d_i$ (initial) and $d_l$ (local) 
with a confidence factor $\lambda \in [0,1]$:
\begin{equation}
d = \lambda d_l + (1-\lambda) d_i .
\label{eq:pers}
\end{equation}
\iftechnicalreport{When $\lambda=0$ we have no trust in the calculated path-specific threshold, and similarly when $\lambda=1$ we have reached a state with full confidence in the path-specific threshold. In practice this should occur when we have seen infinitely many valid paths variations of both our reference path, and we believe no additional individual observation will change the decision threshold.}\fi
We call the resulting threshold $d$ a \emph{mixed threshold}.
A common way to model the confidence in small sample sizes is to use add-one smoothing~\cite{murphy2012machine}. When $n$ is the number of seen instances of a reference path, we can model $\lambda$ as:

\begin{equation}
\lambda(n) = \dfrac{n-1}{n}.
\label{eq:lam}
\end{equation}
This fulfills our boundary conditions for $\lambda$: $\lambda(1) = 0$, implies no confidence when we have only seen one instance of a reference path and $\lambda(\infty) = 1$, signifying full confidence with infinitely many instances.
Figure~\ref{fig:cf} shows how the confidence factor increases w.r.t. the number of reference path instances seen so far.
We use equations~\ref{eq:pers} and~\ref{eq:lam} to determine mixed thresholds throughout this paper. The three thresholds mixed, local and initial are evaluated in Section~\ref{sec:several}.

\iftechnicalreport{
\section{System Design} \label{sec:systemDesign}

In this section we critically evaluate a system proposed in~\ref{sec:systemModel}. We look into common difficulties in designing such a system and propose solutions for overcoming them.

\subsection{A prelude}

One of the difficulties in relying only on on-board sensors is that it is impossible to establish an accurate estimate of the device position based on external coordinate system, since we inherently cannot rely on that external information. The coordinates within a local coordinate system too, are difficult to establish since we rely on noisy sensors: the device coordinate will drift over time, the more the longer recordings we have.

It is known to be difficult to accurately estimate the end coordinate only using the internal-state sensors, and more so, when we require that the end coordinate should be the same each time we use the system.

Modern smart phone sensors noise errors mostly occur due to (device-specific) constant bias and high-frequency white noise~\cite{woodman2007introduction}.
Sensor noise resilience obviously needs to be incorporated into the system from the start.
While accurately pinpointing the location of a device the map might be difficult, we can describe the round-about set of operations (the trajectory) required to reach that point. A one-dimensional sequence of events is more resilient to sensor noise than a two-dimension (or three-dimensional) coordinate: one missed turn will affect the resulting sequence less than what the resulting coordinate difference would indicate. We therefore choose to investigate similarities between trajectories.

There are several ways of representing trajectories and calculating similarities between them. Dynamic Time Warping (DTW) can be used to calculate the similarity between time series, where the continuous values are spaced at fixed temporal intervals. Trajectory analysis for vehicles based on DTW has been conducted by Nawaz et al~\ref{nawaz2014mining} in the past based on gyroscope data alone. We propose an alternate event-based model. In an event-based model, it is essential to recognize events as consistently as possible, i.e. practically always recognize a certain event when it occurs. This limits the possible events we might use for trajectory generation with the device.
We chose to investigate the usage of two main sources of events: turns during a trajectory and movement between the turns.
There are a number of benefits to our model: we can directly incorporate accelerometer-based movement as a symbol, we can represent the trajectory consistently with a small amount of symbols, we can establish and entropy measure for our path and most importantly, we can emulate this path directly with map-based data for a given city, to get a feeling for the actual \textit{uniqueness} of the path within that locality.
The event-based system also makes it possible to incorporate several types of events into a trajectory.
DTW is not appropriate for categorical data; we use similar dynamic programming-based sequence matching algorithms.

To evaluate the security of using the trajectory (rather than coordinate) as an authentication token is the principal result for this paper. We evaluate how much sensory error such recordings will have, and establish a predefined decision threshold based on this that gives sufficiently low false reject rates. We then evaluate the security of the token on a city-wide scale: how many other trajectories are there within a city that produce a sufficiently similar representation.

Since our method is a trajectory-based system, rather than a location-based system, the application areas a somewhat different. See Section~\ref{sec:discussion} for a discussion on a few proposed application scenarios. Next we will discuss the overall system. The system has essentially three parts: 1) event generation, 2) path comparison and 3) path updates.
}

\subsection{Event generation}

We consider a simplistic, general model for representing movement. We detect the events \textit{M: movement}, \textit{S: stationary}, \textit{L: left turn} and \textit{R: right turn}. The movement mode (M/S) is classified continuously, while turn events are produced as they occur. Turns are represented as discretized actual angles (a fixed number of angles).
Any movement type is accepted (e.g. driving, bicycling or walking). The list of events, and the simplified modules responsible for generating them can be viewed in Figure~\ref{fig:ed1}.
We remove $S$ symbols from a path before any comparison between two trajectories is done.
The time that each event was generated is logged to the system, and used to automatically trim compared strings to the same temporal length.
We discuss usage of several movement modes in Section~\ref{sec:discussion}.

\begin{figure}[tbh]
    \centering
    \includegraphics[width=0.55\columnwidth]{pictures/event_detection.png}
    \caption{A simplified view of the event detection system. We detect movement \textit{M}, non-movement \textit{S} and turns \textit{L, R} in this paper. }
    \label{fig:ed1}
\end{figure}





\subsection{Candidate path recognition}

To recognize a candidate path we need to calculate the similarity of it w.r.t. the reference path. Both paths can be represented with strings, so we will naturally calculate the string similarity of paths (edit distances). The Levenshtein distance represents the minimum number of \textit{inserts}, \textit{replace} and \textit{delete} operations necessary to change one string $a$ into another string $b$. The Damerau-Levenshtein distance additionally allows \textit{transposition} of adjacent characters. Both algorithms are commonly used in spellcheckers~\cite{}\added[id=MJ,remark={add citations}]{}. The Needleman-Wunsch and Smith-Waterman algorithms are respectively the global and local alignment algorithms in bioinformatics. Matching correct  symbols incur a \textit{reward}, while \textit{inserting}, \textit{replacing} or \textit{deleting} symbols incurs a penalty, \textit{in addition to missing the reward}. The algorithms measure similarity between two strings $a$ and $b$, and in contrast to the edit distances, large values are favorable~\cite{durbin1998biological}. Biological sequences are often long, and use a limited alphabet, similar to our events in Figure~\ref{fig:ed1}. All algorithms differentiate between strings based on the sequences in which operations occurred. Given that we always know the reference path that we compare against, we can count the number of $M$ symbols in it, and do comparison only for candidate paths with the same number of move symbols. No $M$ symbols are generated during turns. For ease of comprehension, we chose the edit distance as our similarity measure, however with adapting weights most suitable for our comparison.

\subsection{Multipath learning: path updating}

In practice, the user will gather more examples of his path as he continues using the system. For security reasons, each new path example needs to be verified by the user. For usability reasons, the system should be operational with only one seen example of the reference path and more available examples should ideally only increase the usability of the system (since the security of the system should be high already with one path).


%
The previous two subsections have discussed how to recognize whether one path looks similar to another path, only by pairwise comparisons between two individual examples of reference paths. 
If we receive several examples of the reference path, we can choose the best representative among those paths. Multipath learning can be done iteratively as more reference paths are traversed. Ideally, the use of a good representative should lower the FRR without increasing the FAR, while still be computationally scalable. The representative can be chosen by analyzing the seen examples of that reference path: the within-class distances.

With only one path, a preset \emph{global threshold} will be used to determine whether or not a candidate path is similar to the reference path. With more paths, the within-class distances can be used to adjust the decision threshold. As the user walks his paths several times, some variation in the paths will occur, and ideally we would like to use this variation to either slide the threshold forward (more permissive) or backwards (less variation). We can calculate a new threshold, a \textit{local threshold}, using the seen examples. To create a new threshold, it is necessary to have both good examples (within-class similarities) and bad examples (between-class similarities). The between-class similarities should be representative of real paths, e.g. simulated paths using a first-order \textit{Markov chain} or random paths picked from a map. A Markov chain is often represented with a transition matrix, representing the transition probabilities to all possible next states (including itself).

However, using only the local threshold exposes us to problems of using extremely small sample sizes. Therefore, ideally we would like to initially not trust the local threshold, but gradually trust the local threshold more, once we have more examples of the reference path. A fused, \textit{mixed threshold} can be derived, which initially starts at the global threshold and gently slides towards the local threshold as we gain more examples of that reference path.
The next section explains how we practically implemented the proposed systems in this section.

\else{

}
\fi

\section{Implementation and Evaluation} \label{sec:experimentalEvaluation}

\begin{figure}[tbh]

\centering
\begin{minipage}{0.4\columnwidth}
\centering
\includegraphics[width=.9\textwidth]{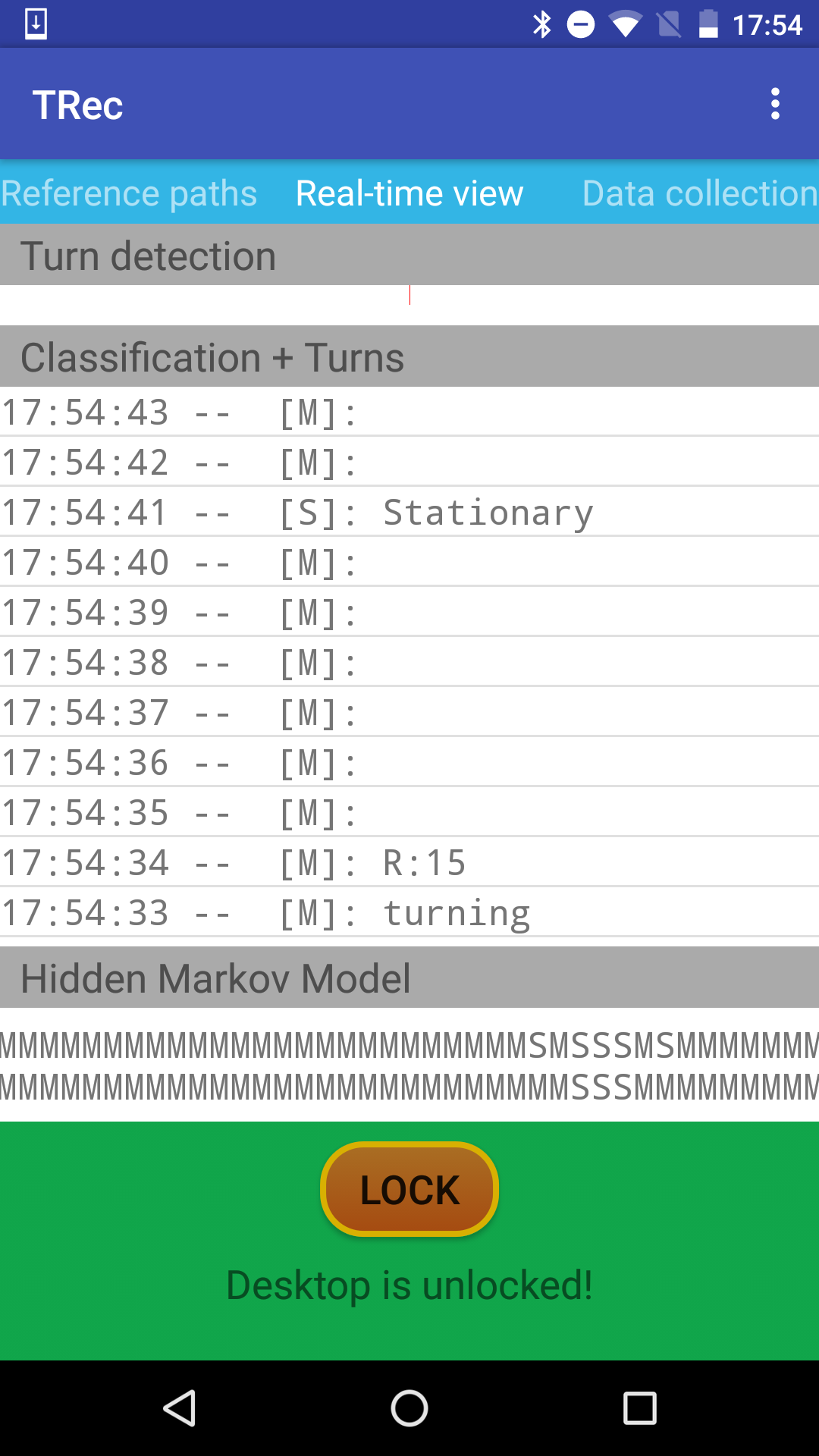}
\end{minipage}
\begin{minipage}{0.4\columnwidth}
\centering
\includegraphics[width=.9\textwidth]{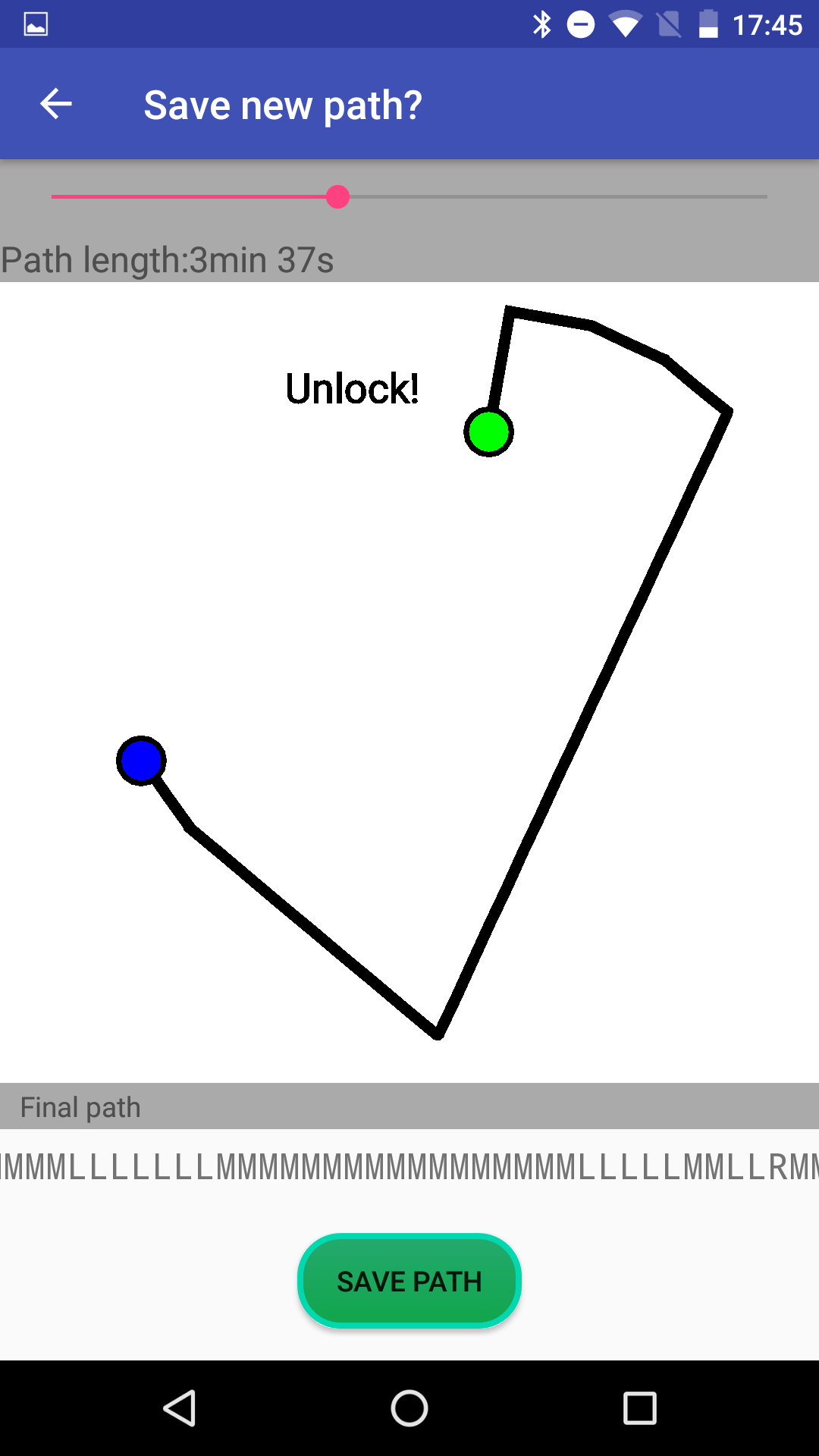}
\end{minipage}
\caption{
\sysname classifies movement at one second intervals. \emph{Left:} Users can quickly lock \ver in case of false accepts. \emph{Right:} Interface for saving unrecognized paths. The length of the reference path can be adjusted.}
\label{fig:android}
\end{figure}

To be able to evaluate the performance and resource requirements of \sysname under real-world conditions, we implemented our system on Android (See Figure~\ref{fig:android}).

\begin{figure*}[tp]
  \begin{minipage}{0.32\textwidth}
    \centering
    \includegraphics[width=.9\textwidth]{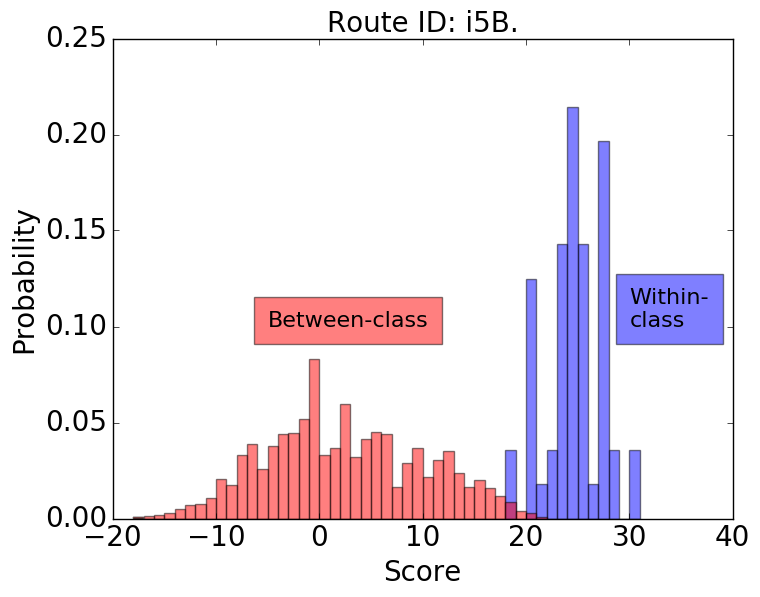}
    \caption{
      An example of within-class (blue) and between-class similarities (red) for one route. Choosing a reference path-specific threshold 18 minimizes equal error rate (EER) on this route.}
    \label{fig:dist}
  \end{minipage}
  \begin{minipage}{0.05\textwidth}
  \end{minipage}
  \begin{minipage}{0.32\textwidth}
    \centering
    \includegraphics[width=.9\columnwidth]{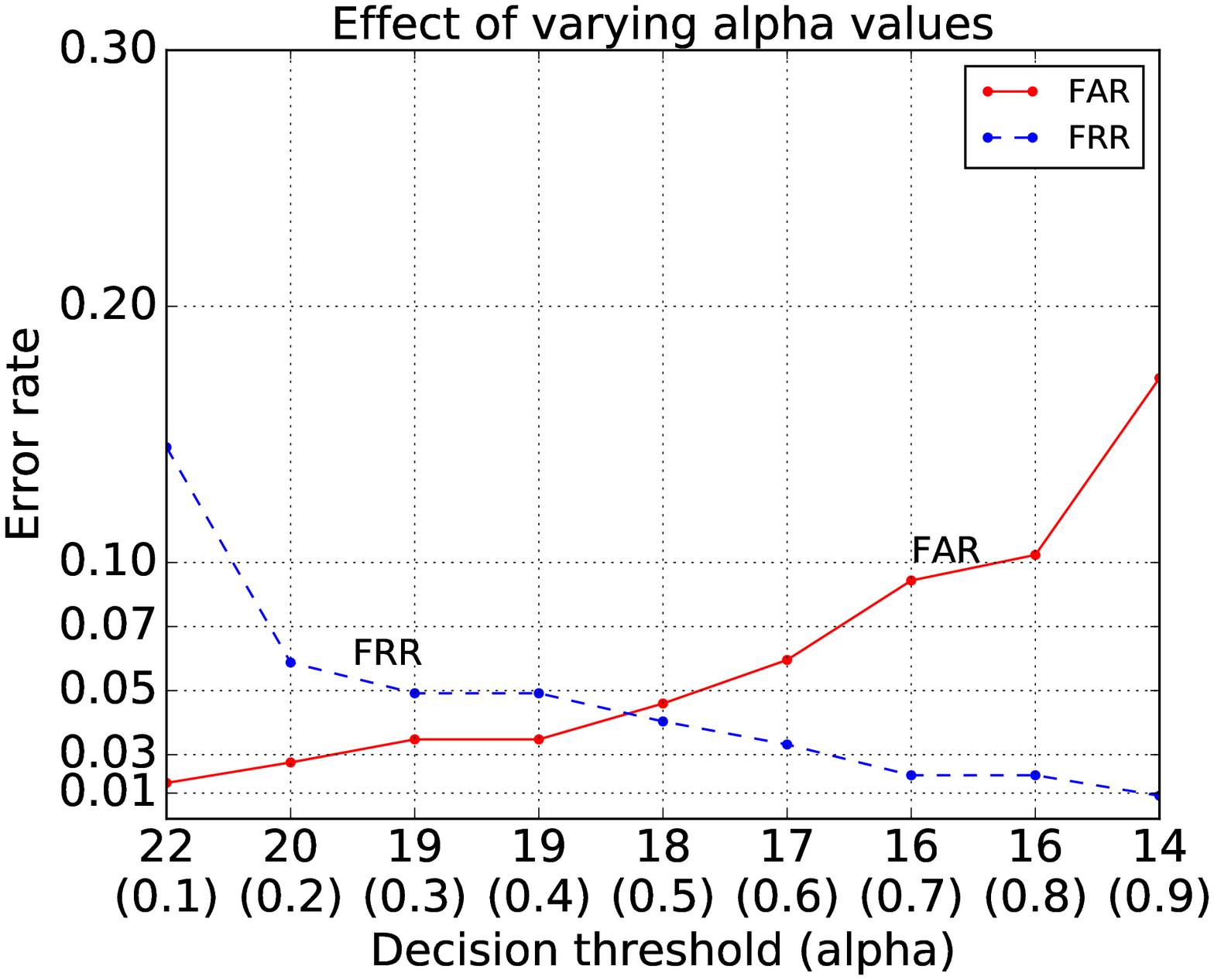}
    \caption{Error rates vs. $\alpha$ (with corresponding decision threshold). Large $\alpha$ values result in low FRR, while small $\alpha$ results in low FAR.
} 
    \label{fig:alphas}
  \end{minipage}\hfill
  \begin{minipage}{0.32\textwidth}
    \centering
    \includegraphics[width=.95\columnwidth]{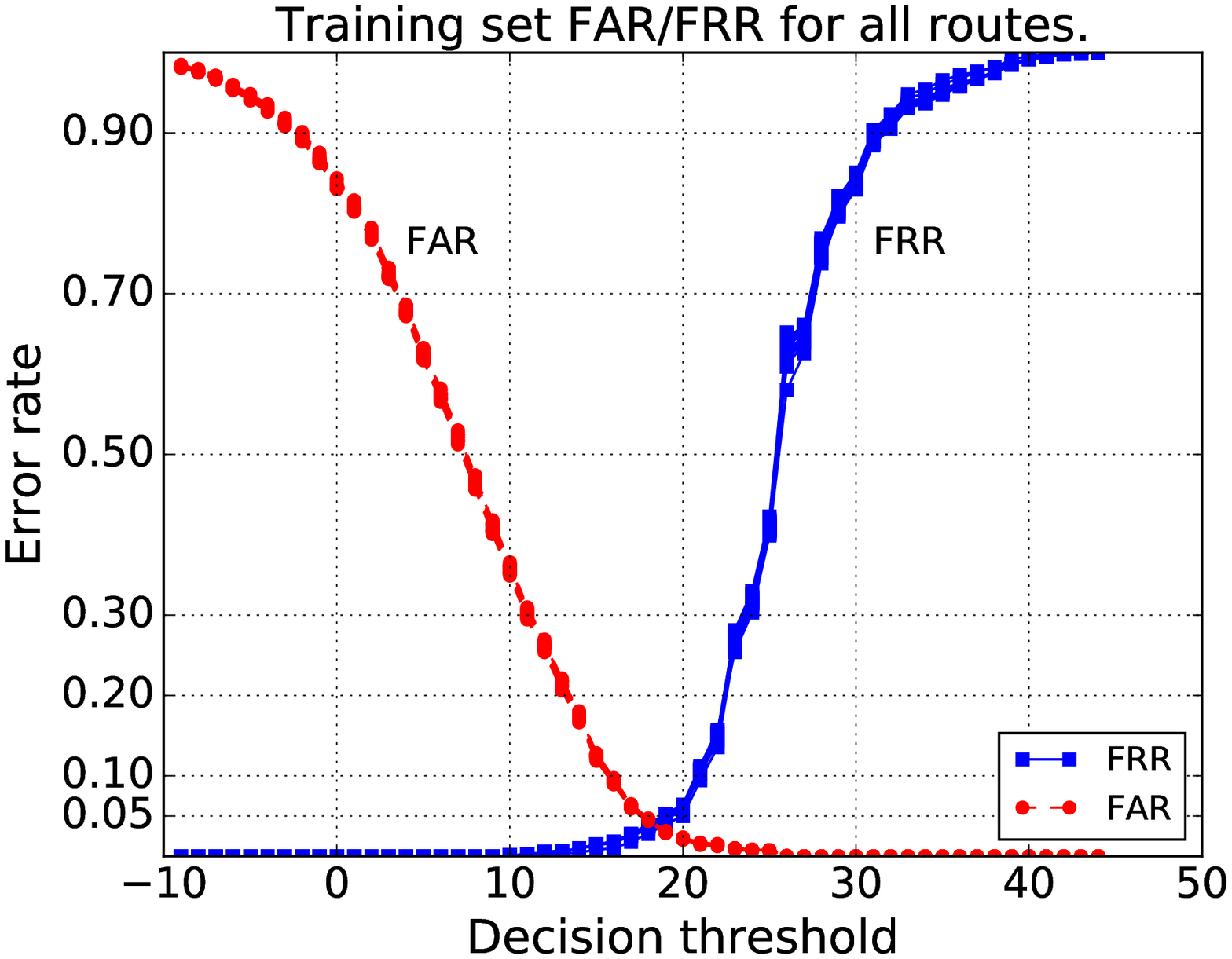}
    \caption{Training set FAR (red) and FRR (blue) for 20 routes (20 graphs). Training set for each route consists of paths of \emph{all other routes} pooled together. EER is achieved at 18 in all graphs.} 
    \label{fig:farfrr}
  \end{minipage}\hfill
\end{figure*}

\subsection{Prototype for \sysname}
\label{sec:req}

To optimize power consumption, the device enters sleep after being stationary for 5 minutes. \sysname acquires a wakelock~\cite{wakelock} when significant motion is detected to ensure that no relevant sensor data is lost due to power optimization.

\myparagraph{Resource requirements}
\sysname expects a continuous stream of raw accelerometer and gyroscope data to extract the current trajectory. It re-samples the input data to 20Hz before storing it into a memory buffer (2.3MB/h each with 32bit precision), which is a fixed circular buffer holding the past hour of measurements. The values are classified on-demand with LR and the classification results are stored into a separate circular buffer (3.6kB/h with 8 bit booleans).
By calculating gravity and linear accelerometer values from raw accelerometer data, \sysname's memory buffer requirements are brought down to less than 5MB in total.
We use weka~\cite{hall2009weka} to model logistic regression on Android.
The data is only smoothed when trajectory comparison is needed. Once triggered, comparison is done continuously at one second intervals for a set of ten attempts by default. If successful, the data is written to disk and a response is generated. Otherwise the authentication attempt is aborted, whereby explicit proximity confirmation is required.

\subsection{Movement Recognition}

To evaluate the accuracy of movement recognition, we collected a
preliminary dataset covering different motion conditions over four hours.
We applied noise-based regularization \cite{murphy2012machine} to reduce potential over-learning.
We then evaluated the resulting LR model using five-fold stratified cross-validation to separate training and testing data.

Out of ten features, we found that the three most significant features were the standard deviation of the 5 second and 1 second sliding window of 3D differentiated accelerometer values, and the peak-to-peak value for the 5 second sliding window of gyroscope measurements: these corresponded to 58\% of the weights of the LR model.
Evaluation results for LR showed a true positive rate of 98\% for movement (M), and 92\% for stationary (S). 
These probabilities are used as emission probabilities in the HMM, and we use a default value of 99\% probability to switch between hidden states.

\subsection{Experimental Data Acquisition}

To evaluate the accuracy of \sysname as a whole, we collected real-world data by repeatedly traversing a series of routes.
The dataset consists of paths corresponding to 20 different, 6 to 12 minute long routes in \ifanonymous{two cities}\else{Espoo and Oxford}\fi.
In order to generalize across different devices, we gathered data from five different device models\footnote{OnePlus One, Nexus 6, Moto G (gen 2), Nexus 5X and Samsung GS6.} at 200Hz sampling rate.
We integrated the measurement devices with bicycles to avoid uncontrolled orientation changes.
Each trajectory was repeated between 7 and 11 times.
Routes contain real-world obstacles, such as traffic lights, gravel, asphalt or cobblestone roads and crowds.
In total, the 7.7 GB dataset consists of the equivalent of 38.6 hours of recording, collected over a total distance of 123 km.

\subsection{Path Similarity Thresholds}
\label{subsec:thresholds}
\myparagraph{Separability}
For each of 20 unique routes $R_k, k \in\{1,\dots,20\}$, our data contains between 7 and 11 instances $i_{R_k}$.
In order to evaluate the within-class similarities of each route and determine if the Needleman-Wunsch measure supports consistent classification, we calculated the similarities between all pairs of instances, trimmed to a duration of 2 minutes. 
For each route, this results in at least 21 unique pairwise within-class, and 1595 unique pairwise between-class, similarities.

Figure~\ref{fig:dist} shows an example of within-class and between-class similarities for one such route in our dataset, with all other routes showing similar behavior.
In an ideal noiseless case, all instances $i_R$ of a route $R$ are identical, but in real-world data, sensor noise results in a spread of similarities.
While there is an overlap of within-class and between-class similarities in the score range [18,21], most within-class and between-class cases are separated, confirming that Needleman-Wunsch is indeed a good measure of similarity.

\myparagraph{Determining the initial threshold}
As described in Section~\ref{sec:implementation}, decision thresholds that \sysname uses are adapted based on the number of instances of a reference path seen so far.
When only a single instance has been authorized by the user, \sysname uses the \emph{initial threshold}; as more instances are seen, their similarities to the reference path are used to compute a \emph{local} threshold, and subsequently the \emph{mixed} threshold, which is a combination of initial and local thresholds that depends increasingly less on the initial threshold as the number of seen instances of a reference path increases.

We can compute the threshold with respect to a chosen trade-off between FAR and FRR by minimizing the combined error rate $\alpha \cdot FRR + (1-\alpha) \cdot FAR$ for a specific value of the trade-off parameter $\alpha$.
Figure~\ref{fig:alphas} shows optimal achievable FARs and FRRs for pooled\footnote{We aggregated all reference-path specific within-class distances to one group, and all reference-path specific between-class distances to another group.} reference paths.
For each value of $\alpha$, the minimum combined FAR/FRR is found by varying the decision threshold.
In scenarios where the usability of the system is more important, larger $\alpha$ values should be used.

The decision threshold is naturally tied to the length $L$ of the reference path. 
In order for \sysname to make a classification decision for a candidate path given only one instance of a reference path, we need to determine a function for the initial threshold that depends on the length of the reference path $L$.
We did this by searching for the value $d_L$ that minimizes the combined error rate for $\alpha \in \{0.1, 0.2, \dots, 0.9\}$ for reference path lengths $L \in \{1,2,\dots,6\}$.
This gave us 6 dependent-independent variable pairs per $\alpha$, i.e. 9 linear regressions \cite{murphy2012machine}.
We found that the relationship between optimal decision threshold and the length of the path is affine. Figure~\ref{fig:alphas} shows that $\alpha^*=0.5$ is closest to equal error rate.
Setting  $\alpha^*$ to 0.5, we obtained the decision thresholds for pooled scores as:
\begin{equation}
\begin{split}
D^*(L) &= 9.69L - 1.40
\end{split}
\label{eq:decision_thresholds}
\end{equation}

\noindent
These are rounded to integers such as: $D^*(1) = 8$, $D^*(2) = 18$ and $D^*(5) = 47$. These particular decision thresholds serve as examples on how initial thresholds are calculated, and we use them in our \sysname prototype.
However, in order to give an unbiased estimate of system performance, we do not use exactly these values in the remainder of system evaluation, as this might lead to over-fitting.

\begin{table}[tp]
\caption{Mean and standard deviation of FAR/FRR using only the \textbf{initial decision threshold} individually for 20 paths.}
\begin{center}
  \begin{tabular}{| l | c | c |}
    \hline
$L$ (min) &FAR &FRR  \\ \hline \hline
1.0 min & $\textbf{0.064}\pm0.057$ & $\textbf{0.137}\pm{0.143}$ \\
2.0 min & $\textbf{0.043}\pm0.048$ & $\textbf{0.039}\pm{0.102}$ \\
3.0 min & $\textbf{0.073}\pm0.085$ & $\textbf{0.047}\pm{0.107}$ \\
4.0 min & $\textbf{0.072}\pm0.074$ & $\textbf{0.061}\pm{0.133}$ \\
5.0 min & $\textbf{0.049}\pm0.061$ & $\textbf{0.057}\pm{0.103}$ \\
6.0 min & $\textbf{0.047}\pm0.070$ & $\textbf{0.040}\pm{0.081}$ \\

\hline
\end{tabular}
\end{center}
\label{table:global_path}
\vspace*{-7mm}
\end{table}

Instead, we calculate decision thresholds for each reference path separately, using the \emph{leave-one-reference-path-out} method, which we use for all subsequent analysis (a training set is constructed using 19 reference paths; it is used to determine the initial threshold for the remaining reference path).
Figure~\ref{fig:farfrr} shows the FARs and FRRs by removing one and retaining the 19 other reference paths (there are 20 lines in total).
As can be seen, training set error rates below 5\% are achievable for all 20 sets.

The next two sections discuss actual test error rates for specific paths that use individual and mixed thresholds, evaluated w.r.t. two different parameters: the reference path length and the number of instances of a reference path.

\vskip-0.05cm\noindent
\subsection{Impact of Reference Path Length}
The mean error rates and standard deviations for our 20 reference paths using the initial threshold with varying reference path lengths are shown in Table~\ref{table:global_path}.
FRRs are calculated over all unique combinations of dividing instances $i_R$ of routes $R$ into reference path instances (training) and candidate paths (testing).
FARs are calculated by treating all other instances $i_{\bar{R}}$ of routes $\bar{R}$ as candidate paths.
We know the temporal length of the reference path and require the same length for each candidate path.
Both FAR and FRR drop when increasing the length of the path from one to two minute, and reach lowest mean FAR and FRR on 2 and 6 min long reference paths. 
Initially, the mean FAR is between \textbf{4.3\%} and \textbf{7.3\%} and FRR between \textbf{3.9\%} and \textbf{6.1\%} for reference paths longer than two minutes. However, the variation in FRRs is high between different reference paths.
We see a clear drop at 2 min, but FARs/FRRs for individual paths do not change significantly beyond $L = 2$ min (Wilcoxon signed rank test\footnote{Null hypothesis: No change in median FAR (resp. FRR) by increasing paths' temporal length. FAR: p-value 0.47, FRR: p-value 0.64. Wilcox zero treatment, sample size = 79, test statistics 182.5 and 1023.}).
We thus use $L = 2$ min for the rest of the analysis in this paper.

\vskip-0.05cm\noindent
\subsection{Using Multiple Reference Path Instances}
\label{sec:several}
Multiple instances of a reference path can be used to increase security and usability of the system.
This is done by selecting a good representative for the reference path among the seen instances (the instance at the cluster center - the medoid) and by adopting a mixed threshold using equations~\ref{eq:pers} and~\ref{eq:lam} (combination of $d_i$ and $d_l$). The local decision threshold is calculated with the same scheme as the initial decision threshold, by setting $\alpha=0.5$.
If there are multiple solutions that provide the same minimum error rate, we use the decision threshold that is center most among these for a maximum margin.
The difference to the initial threshold derivation is that we generate paths to estimate the distribution of between-class similarities using a Markov chain~\cite{murphy2012machine}. This is done to prevent over-learning by only using collected, real world paths.

\begin{figure}[tp]
\centering
\includegraphics[width=.9\columnwidth]{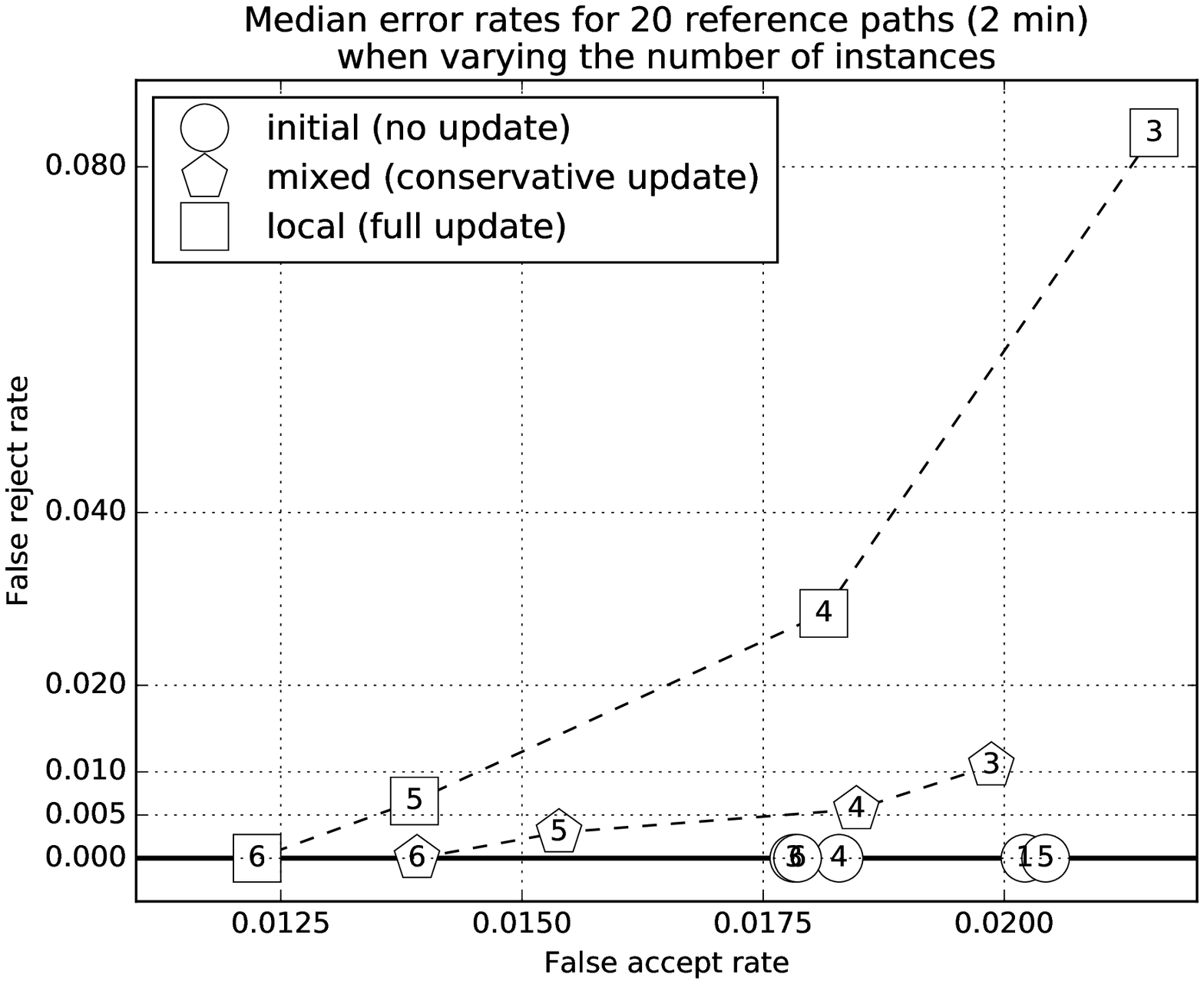}
\caption{
Increasing the number of instances of reference paths leads to a median decrease in FAR and FRR. The extent of this effect varies depending on the used threshold derivation algorithm. Coordinates describe median (FAR, FRR), numbers describe number of instances used.
}
\label{fig:n_paths}
\vspace*{-7mm}
\end{figure}

To evaluate the effects of using the medoid and Equation~\ref{eq:lam} to calculate the confidence factor $\lambda$ we plotted the median FARs and FRRs (per reference path), by varying the number of instances and the decision threshold calculation scheme. It is easier to investigate trends with the median, since it is a robust estimator that tolerates outliers better than the mean.
Figure~\ref{fig:n_paths} shows the median FARs and FRRs for the initial, local and mixed thresholds using 2 min reference paths.
These represent doing no update, doing a full update or a conservative update to the decision threshold, respectively.
At each update, the pairwise similarities are calculated between the
instances, and the instance with the largest summed similarity to the other instances is selected as the medoid. The medoid is used for calculating the similarity to each new candidate path. The benefit in using the medoid to represent the reference path is that only one similarity score needs to be calculated, and that the score is calculated on the most representative instance. 
In Figure~\ref{fig:n_paths}, circles represent the continued usage of the initial threshold.
The median FRR is zero, but the FAR does not improve with more instances, because the FAR depends on the decision threshold, not on the reference path representative.
Using the initial threshold is equivalent to a constant confidence factor $\lambda = 0$, no trust in $d_l$.

The squares represents the effect of using only the local threshold.
While FARs are at a similar level as the initial threshold,
median FRRs are significantly worse. With five or more instances, median FARs drop to a lower level than with initial thresholds. 
Using only the local threshold is equivalent to a constant confidence factor $\lambda = 1$ (full trust in $d_l$).

Mixed thresholds are derived from initial and local threshold values using equations~\ref{eq:pers} and~\ref{eq:lam}, and are shown with pentagons.
The mixed thresholds retain the good FRR of initial thresholds when few reference path instances are seen, and achieve improved performance similar to local thresholds when more instances have been observed.
Using the confidence factor in conjunction with medoids to calculate decision thresholds is empirically shown to increase performance on two-minute paths, dropping the median FAR from $2.0\% \to 1.5\%$ when increasing the number of instances from 1 to 5, while the median FRR increases from $0.0\% \to 0.3\%$. Simultaneously, the mean FAR drops from $4.3\% \to 3.4\%$ and mean FRR drops from $3.8\% \to 2.8\%$. To further validate our results, we analyze the results of \sysname when \textbf{five instances} are used with \textbf{mixed thresholds}.

Table~\ref{table:mixed_path} shows the resulting mean FARs and FRRs for different reference path lengths. Mean FARs are in the range of \textbf{2.3\%} to \textbf{5.0\%} and mean FRRs are in the range of \textbf{1.8\%} to \textbf{3.1\%} for reference paths longer than 2 min. Both FAR and FRR are smaller for these paths, and the spreads in FRRs are significantly lower than in Table~\ref{table:global_path}. We find again that FARs/FRRs for individual paths do not change significantly by considering reference paths beyond 2 min (Wilcoxon signed rank test\footnote{Null hypothesis: No change in median FAR (resp. FRR) by increasing paths' temporal length. FAR: p-value 0.43, FRR: p-value 0.15. Wilcox zero treatment, sample size = 79, test statistics 779.5 and 272.}).

\begin{figure}[tp]
\centering
\includegraphics[width=.96\columnwidth]{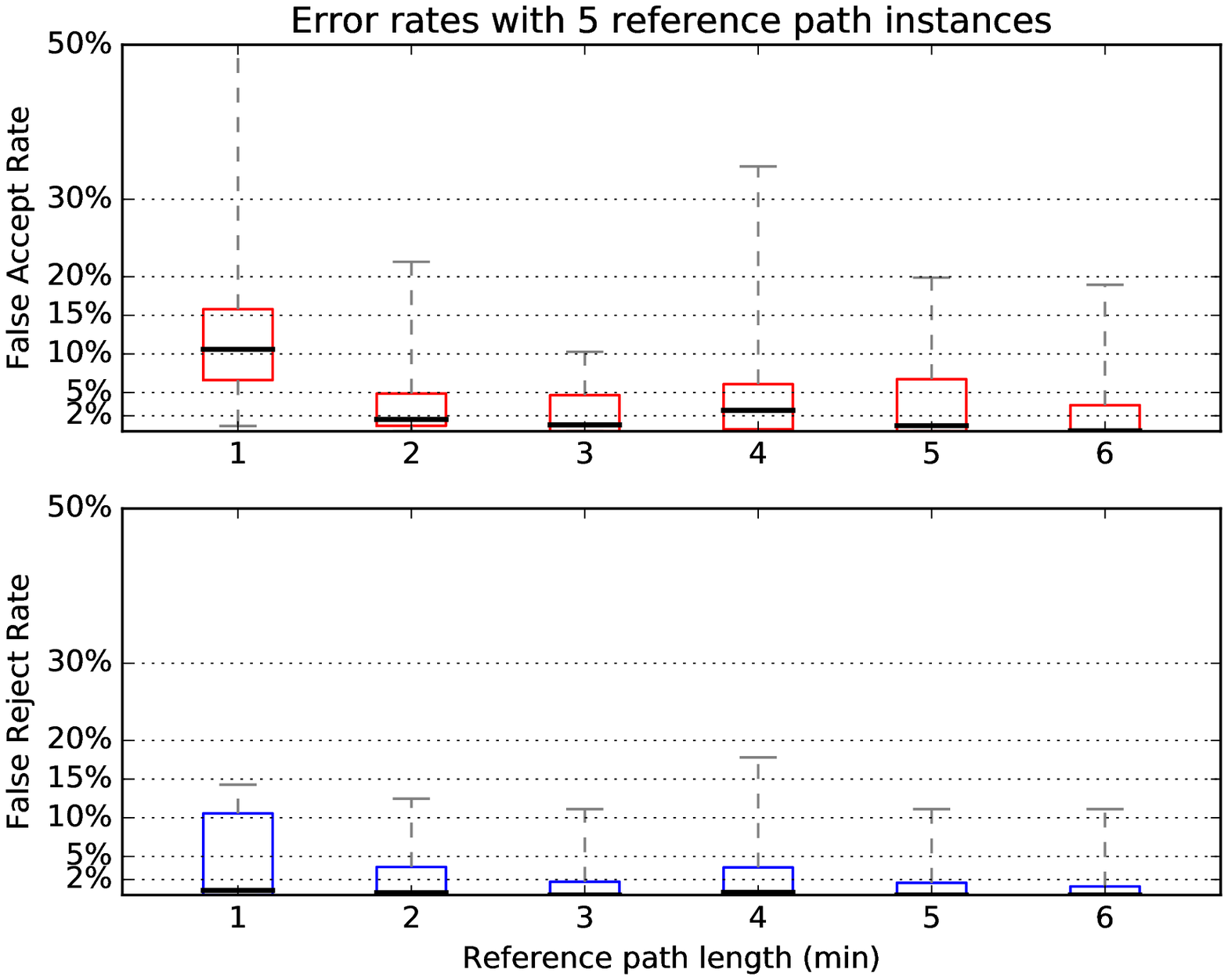}
\caption[width=.96\columnwidth]{\textbf{Mixed decision thresholds}. Box-and-whiskers plot over FARs and FRRs of different temporal lengths when \sysname has seen \textbf{five instances} of a reference path. On average FARs are below 3\% and FRRs are below 0.5\% when $L \ge 2 min$.}
\label{fig:len_paths_5}
\vspace*{-5mm}
\end{figure}

Figure~\ref{fig:len_paths_5} shows box-and-whiskers plots~\cite{murphy2012machine} for FARs and FRRs for our 20 reference paths, evaluated over different lengths, once five instances have been observed by \sysname.
The outer lines of the boxes denote the 75th and 25th percentiles, while the line in the boxes denotes the median. The whiskers denote the range, i.e. maximum and minimum values.
Note that on average, FARs and FRRs are lower than the mean values reported in Table~\ref{table:mixed_path}.
\begin{table}[h]
\caption{Mean and standard deviation of FAR/FRR using \textbf{five instances} and \textbf{mixed decision thresholds} for 20 paths.}
\begin{center}
  \begin{tabular}{| l | c | c |}
    \hline
$L$ (min) &FAR &FRR  \\ \hline \hline
1.0 min & $\textbf{0.141}\pm0.135$ & $\textbf{0.046}\pm{0.054}$ \\
2.0 min & $\textbf{0.034}\pm0.049$ & $\textbf{0.028}\pm{0.042}$ \\
3.0 min & $\textbf{0.023}\pm0.028$ & $\textbf{0.018}\pm{0.031}$ \\
4.0 min & $\textbf{0.050}\pm0.078$ & $\textbf{0.031}\pm{0.050}$ \\
5.0 min & $\textbf{0.042}\pm0.058$ & $\textbf{0.017}\pm{0.032}$ \\
6.0 min & $\textbf{0.030}\pm0.050$ & $\textbf{0.015}\pm{0.032}$ \\
\hline
\end{tabular}
\end{center}
\label{table:mixed_path}
\vspace*{-3mm}
\end{table}
For reference paths longer than 2 min, median FARs are below 3\% and FRRs below 0.5\%.

Our implementation of \sysname uses mixed thresholds so that its performance improves with time.

\subsection{Energy Consumption}
\label{sec:battery}

To evaluate requirement~R\ref{R-efficiency}, we created a controlled experiment to measure the energy consumption of \sysname. We obtained 3-hour consumption reports on three different devices\footnote{Nexus 5, Nexus 5X and Samsung GS6. The devices are charged to 100\% before the start of the experiment; the battery level is measured after 3 hours.}, ten times each. A control group had no \sysname installed, and did not have WiFi nor mobile connectivity.

\myparagraph{Case 1}
We model a scenario where an office employee periodically moves away from her workplace and returns back, triggering an access control decision. 
Movement initiates data collection at full sampling rate. 
Upon approaching the premise, and thus \ver, path recognition is triggered and runs 60 times.\footnote{The employee moves for 5 minutes once an hour, repeated three times.}
Table~\ref{table:energy} reports the net contribution of \sysname, compared to the control group.\footnote{Android Debug Bridge provides lower and upper bounds to the battery usage with 1\% granularity. We only show the upper bound estimates here.}
The increase in energy consumption varies between devices. On average we recorded a drain of $0.56\%$ (of total battery capacity) per hour.

\begin{table}[h]
\caption{
Net increases in hourly battery consumption rates (percentage of total capacity) and milliampere (case 1).
}
\begin{center}
  \begin{tabular}{| c | c c |}
    \hline
Device & $\%/h$ & $mA$ \\ \hline
Nexus 5 & $0.41\pm0.23$ & ($9.32\pm 5.22$)\\
Nexus 5X & $0.63\pm0.33$ & ($17.13\pm 8.92$)\\
Samsung S6 & $0.61\pm0.24$ & ($15.45\pm 6.44$)\\
\hline
  \end{tabular}
\end{center}
\label{table:energy}
\vspace*{-3mm}
\end{table}

\myparagraph{Case 2} We observed that the largest contributor to battery usage was the wakelock that ensured sensor measurements were processed quickly and not lost due to an overflowing sensor buffer. Therefore, we recorded the energy usage with wakelock acquired for three hours. This represents the worst case scenario. The hourly net battery drains for our devices are shown in Table~\ref{table:energy2}. The mean increase in battery usage varies across devices, with an overall average of $1.9\%$ per hour.
We believe that the drain increase in Nexus 5X and Samsung S6 is a result of these devices using different power states efficiently, causing the consumption in the control group to be lower.

\begin{table}[h]
\caption{
Net increases in hourly battery consumption rates (percentage of total capacity) and milliampere (case 2).
}
\begin{center}
  \begin{tabular}{| c | c c |}
    \hline
Device & $\%/h$ & $mA$  \\ \hline
Nexus 5 & $0.87\pm0.28$ & ($20.11\pm6.40$)\\
Nexus 5X & $2.31\pm0.24$ & ($62.28\pm6.60$)\\
Samsung S6 & $2.55\pm0.17$ & ($65.14\pm4.33$)\\ \hline
  \end{tabular}
\end{center}
\label{table:energy2}
\vspace*{-3mm}
\end{table}

\myparagraph{Case 3}
Energy consumption of smart phone models can vary significantly for different settings and sensors. On average users recharge their batteries within an hour of depletion~\cite{wagner2013device}.
A study~\cite{peltonen2015energy} found the average battery life to be nine hours (11\% consumption rate) under favorable circumstances.
In a setting where \user commutes for one hour and does office work for eight hours, we estimate (using data from Tables~\ref{table:energy} and~\ref{table:energy2}) the additional battery drain due to \sysname to be between 4\% and 7\%, depending on the phone model. These correspond to a decrease of approximately 20-40 minutes for a device with a battery life of nine hours.
For comparison, the loss in battery life when moving from an area with average WiFi to bad coverage was estimated to be 6.29\%~\cite{peltonen2015energy}.

\section{Security Analysis}

\sysname prevents relay attacks by having \pro verify its proximity to \ver through prover-side trajectory comparison:
if \pro's current trajectory (candidate path) matches the authorized approach trajectory (reference path) to \ver, \pro locally decides to participate in the authentication protocol with \ver.

To evaluate the security guarantees of \sysname, we analyze our dataset (Section \ref{sec:experimentalEvaluation}) corresponding to a total of 123~km equivalent to 38.6 hours of movement.
We used false accept rate of trajectory comparison as the measure for security. We showed that, on average, an adversary has a less then 5\% chance even when \sysname has only observed one instance of a reference path (Table~\ref{table:global_path}). This drops steadily as more instances are collected so that we can report a rate below 3.5\% in Table~\ref{table:mixed_path} on average and 1.5\% in median (Figure~\ref{fig:n_paths}) after 5 path instances.

For a successful relay attack (a) \adv is required to follow \user in close proximity and (b) \pro has to be in motion in order to produce a path recording. In a typical attack scenario, a thief tries  to gain access through \ver to steal a keyless-entry car while \user and \pro are stationary which leads to a failing proximity verification and thus \pro denying access to the authentication credentials. In this case \sysname successfully improves security by \textit{completely} preventing the relay attack.

The security of the trajectory leading to \ver is inversely proportional to the likelihood that there exists another route (not terminating in \ver) that \adv could use to mount a relay attack \emph{if} \user happens to traverse it.
The measure for security however depends on the geographic neighborhood: for instance, a trajectory with many turns can be considered to be more complex than a straight-line in most cities but in an old town that has few straight roads, such a trajectory may be uniquer. We can incorporate such uniqueness assessment of a reference path into \sysname by analyzing a city's road data to find paths that are similar to a given reference path and hence vulnerable to relay attacks. This is equivalent to the likelihood of the user unintentionally authenticating to \ver while she is traversing some other route not leading to \ver. This estimate can be used as a security level indicator.

\section{Discussion and Future Work}
\label{sec:discussion}

\myparagraph{Why transparent authentication}
When trajectory comparison fails, \sysname will prompt the user for explicit proximity confirmation as to whether the current candidate path should be included in the set of authorized reference paths for a given verifier. The popularity of keyless entry as a premium feature across different car manufacturers suggest that consumers are willing to pay for the convenience of transparent authentication. Google's Trust API (Project Abacus)~\cite{Hern2016} underscores TA as a trend.  \sysname's contribution is to \emph{retain} the convenience promised by transparent authentication while significantly enhancing resilience against relay attacks.

\myparagraph{Verification of return paths}
By relying on reference path instances gathered over time, \sysname is currently limited to scenarios with stationary verifiers. But since \sysname represents trajectories as sequences of primitives it is possible to compare a reference path that \emph{starts} from the location of the verifier to a candidate path that ends at the verifier by reversing the reference path.
This allows transparent authentication even for \emph{mobile verifiers}: e.g., when a user parks his car at a new place and goes to a mall and returns to the car via the same route. 
Our evaluation showed promising results, and we leave its comprehensive analysis for future work.

\myparagraph{Prover orientation changes}
If \pro is a portable device like a smart phone, fast changes in \pro's orientation (e.g. taking the phone out of the pocket) affect turn detection reliability. In our evaluation, we avoided this by integrating \pro with a vehicle, we chose a bike for our experiments.

\myparagraph{Additional sequence primitives}
While this paper focuses only on using low-cost internal sensors, \sysname can be extended to use additional primitives besides movement, left and right turn.
If the requirement to use only internal sensors is relaxed, the system could start detecting presence of specific wireless networks or indoor short-range beacons; if the user is indeed taking the same path as was the case when the corresponding reference path was recorded, then the same events should be detected at specific locations.
Such detected presences can easily be represented by additional symbols and seamlessly integrated into the current implementation of \sysname.
Additionally, while we currently focus only on a single movement modality, \sysname can be augmented with a transport mode detection scheme~\cite{hemminki2013accelerometer} that would add further entropy and hence uniqueness to generated paths.
As such, a path that includes walking, then taking a bus, and then walking would be more specific than if all movement intervals are represented with the same symbol.

\section{Related Work} \label{sec:relatedWork}

\myparagraph{Trajectory recognition}
Trajectory recognition estimates the location of a user and can be used to enable additional convenience functions but also to subvert a user's privacy. The following techniques do not rely on GPS location information but use a different way to obtain location ground truth, e.g. a street map or the timetable of public transportation. After a correlation between mobile measurements and this information the system estimate the user's location.

Gao et al.~\cite{Gao2014} propose an approach using vehicular speed and the start location to estimate final destination and path of a car. By matching these segments to map data they achieved an accuracy of 500 meters for 24\% of traces in the New Jersey and 26\% of traces in the Seattle area.
Further, Watanabe et al.~\cite{Watanabe2015} identify a user's train trips based on inertial measurements. First, the user's activity is classified into inside a \textit{vehicle}, \textit{walking} and remaining \textit{stationary}. Afterwards, the transition times between the different modes are used to correlate them with timetables. Each train trip is weighted according to its popularity to reduce the number of candidates. Their results show that location detection along train networks is feasible.
The work by Nawaz and Mascolo~\cite{nawaz2014mining} explores the significant transport routes of a user based on gyroscope data. According to their hypothesis, a route exposes a certain signature based on angular momentum. They apply dynamic time warping to account for differences in routes due to traffic conditions. Our system in contrast ignores stationary phases which makes time warping unnecessary.
As shown by Narain et al.~\cite{narain2016inferring}, it is possible to infer routes taken by a user solely based on permission free on-board sensors, e.g. gyroscope. Hence, to protect a user's privacy our approach processes the information locally without the need of a remote service.

\myparagraph{Co-presence verification}
The co-location of devices is an important countermeasure against impersonation and relay attacks. In some cases it is also used as a second factor for authentication.
Although GPS could be used to assert co-location in theory, these signals are not authenticated and thus not trustworthy~\cite{Tippenhauer2011}. A range of alternatives based on context comparison has recently emerged.

Halevi et al.~\cite{Halevi2012} propose co-presence detection based on comparing audio and light. A merchant terminal and mobile phone probe their environments to compare them to assert co-location. They evaluate both modalities separately, and achieve a FAR of 6.5\% and a 5\% FRR for light while reporting a FAR and FRR of 0\% for audio.

A similar approach to mitigate relay attacks is explored by Shrestha et al.~\cite{Shrestha2014} and Truong et al.~\cite{Truong2015}. They use natural environment properties as well as digital signals. Truong et al.~\cite{Truong2015} identify WiFi as the dominating feature with a FAR of 2\% and a FRR of 1\%. In their approach, Shrestha et al. conclude that a modality fusion reduces the FRR of up to 24\% and FAR of up to 33\% of an individual features to 3\% and 6\% respectively. However, in follow-up work they were able to increase the FAR from 3\% to 66\% by manipulating a single modality~\cite{Shrestha151104}. Hence, increasing the number of modalities does not necessarily strengthen security as it depends on the weights machine learning models assign to them. A thorough analysis of these algorithms is required to give sophisticated security guarantees.
Karapanos et al.~\cite{Karapanos2015} use the audio fingerprint of a location as a second factor for authentication. Their threat model assumes a remote attacker who obtained the user's credentials.

\section{Conclusion}

We proposed, implemented and evaluated \sysname, a novel approach to prevent relay attacks in transparent authentication schemes. As \sysname is entirely realized on the prover device it allows easy integration into existing systems while preserving the user's privacy at the same time. The performance of our approach and the negligible resource requirements make it a valuable extension of current TA schemes.

\ifanonymous{}
\else{
\section*{Acknowledgments}
This work was supported by the Academy of Finland ``Contextual Security'' project (274951) and the ``CDT Cyber Security'' fund of the Engineering and Physical Sciences Research Council, United Kingdom.
}\fi

\bibliographystyle{abbrv}
\bibliography{references}

\begin{thebibliography}{10}

\bibitem{wakelock}
Android.
\newblock Keeping the device awake, 2016.
\newblock https://developer.android.com/training/scheduling/wakelock.html.

\bibitem{DBLP:conf/sp/BonneauHOS12}
J.~Bonneau, C.~Herley, P.~C. van Oorschot, and F.~Stajano.
\newblock The quest to replace passwords: {A} framework for comparative
  evaluation of web authentication schemes.
\newblock In {\em {IEEE} Symposium on Security and Privacy, {SP} 2012, 21-23
  May 2012, San Francisco, California, {USA}}, pages 553--567. {IEEE} Computer
  Society, 2012.

\bibitem{DBLP:journals/cacm/BonneauHOS15}
J.~Bonneau, C.~Herley, P.~C. van Oorschot, and F.~Stajano.
\newblock Passwords and the evolution of imperfect authentication.
\newblock {\em Commun. {ACM}}, 58(7):78--87, 2015.

\bibitem{Brands1993}
S.~Brands and D.~Chaum.
\newblock {Distance-Bounding Protocols}.
\newblock In {\em Advances in Cryptology — EUROCRYPT '93}, pages 344--359.
  Springer Berlin Heidelberg, Berlin, Heidelberg, 1993.

\bibitem{Corner2002}
M.~D. Corner and B.~D. Noble.
\newblock {Zero-interaction authentication}.
\newblock In {\em Proceedings of the 8th annual international conference on
  Mobile computing and networking - MobiCom '02}, page~1, New York, New York,
  USA, sep 2002. ACM Press.

\bibitem{dolev1983security}
D.~Dolev and A.~C. Yao.
\newblock On the security of public key protocols.
\newblock {\em Information Theory, IEEE Transactions on}, 29(2):198--208, 1983.

\bibitem{durbin1998biological}
R.~Durbin, S.~R. Eddy, A.~Krogh, and G.~Mitchison.
\newblock {\em Biological sequence analysis: probabilistic models of proteins
  and nucleic acids}.
\newblock Cambridge university press, 1998.

\bibitem{Gao2014}
B.~Firner, S.~Sugrim, Y.~Yang, and J.~Lindqvist.
\newblock {Elastic pathing: Your speed is enough to track you}.
\newblock {\em arXiv preprint arXiv:1401.0052}, pages 975--986, 2013.

\bibitem{Francillon2011}
A.~Francillon, B.~Danev, and S.~Capkun.
\newblock {Relay Attacks on Passive Keyless Entry and Start Systems in Modern
  Cars}.
\newblock {\em Network and Distributed System Security Symposium}, pages
  431--439, 2011.

\bibitem{Francis2012}
L.~Francis, G.~Hancke, K.~Mayes, and K.~Markantonakis.
\newblock {Practical relay attack on contactless transactions by using NFC
  mobile phones}.
\newblock In {\em Cryptology and Information Security Series}, volume~8, pages
  21--32, 2012.

\bibitem{Halevi2012}
T.~Halevi, D.~Ma, N.~Saxena, and T.~Xiang.
\newblock {Secure Proximity Detection for NFC Devices Based on Ambient Sensor
  Data}.
\newblock pages 379--396. Springer Berlin Heidelberg, 2012.

\bibitem{hall2009weka}
M.~Hall, E.~Frank, G.~Holmes, B.~Pfahringer, P.~Reutemann, and I.~H. Witten.
\newblock The weka data mining software: an update.
\newblock {\em ACM SIGKDD explorations newsletter}, 11(1):10--18, 2009.

\bibitem{hemminki2013accelerometer}
S.~Hemminki, P.~Nurmi, and S.~Tarkoma.
\newblock Accelerometer-based transportation mode detection on smartphones.
\newblock In {\em Proceedings of the 11th ACM Conference on Embedded Networked
  Sensor Systems}, page~13. ACM, 2013.

\bibitem{Hern2016}
A.~Hern.
\newblock {Google aims to kill passwords by the end of this year | Technology |
  The Guardian}, 2016.

\bibitem{jimenez2009comparison}
A.~R. Jimenez, F.~Seco, C.~Prieto, and J.~Guevara.
\newblock A comparison of pedestrian dead-reckoning algorithms using a low-cost
  mems imu.
\newblock In {\em Intelligent Signal Processing, 2009. WISP 2009. IEEE
  International Symposium on}, pages 37--42. IEEE, 2009.

\bibitem{Karapanos2015}
N.~Karapanos, C.~Marforio, C.~Soriente, and S.~Capkun.
\newblock {Sound-Proof: Usable Two-Factor Authentication Based on Ambient
  Sound}.
\newblock In {\em 24th USENIX Security Symposium (USENIX Security 15)}, pages
  483--498, 2015.

\bibitem{murphy2012machine}
K.~P. Murphy.
\newblock {\em Machine learning: a probabilistic perspective}.
\newblock MIT press, 2012.

\bibitem{narain2016inferring}
S.~Narain, T.~D. Vo-Huu, K.~Block, and G.~Noubir.
\newblock Inferring user routes and locations using zero-permission mobile
  sensors.
\newblock 2016.

\bibitem{nawaz2014mining}
S.~Nawaz and C.~Mascolo.
\newblock Mining users' significant driving routes with low-power sensors.
\newblock In {\em Proceedings of the 12th ACM Conference on Embedded Network
  Sensor Systems}, pages 236--250. ACM, 2014.

\bibitem{Patel2016}
V.~M. Patel, R.~Chellappa, D.~Chandra, and B.~Barbello.
\newblock {Continuous User Authentication on Mobile Devices: Recent progress
  and remaining challenges}.
\newblock {\em IEEE Signal Processing Magazine}, 33(4):49--61, jul 2016.

\bibitem{peltonen2015energy}
E.~Peltonen, E.~Lagerspetz, P.~Nurmi, and S.~Tarkoma.
\newblock Energy modeling of system settings: A crowdsourced approach.
\newblock In {\em Pervasive Computing and Communications (PerCom), 2015 IEEE
  International Conference on}, pages 37--45. IEEE, 2015.

\bibitem{Shrestha2014}
B.~Shrestha, N.~Saxena, H.~T.~T. Truong, and N.~Asokan.
\newblock {Drone to the Rescue: Relay-Resilient Authentication using Ambient
  Multi-sensing}.
\newblock pages 349--364. Springer Berlin Heidelberg, 2014.

\bibitem{Shrestha151104}
B.~Shrestha, N.~Saxena, H.~T.~T. Truong, and N.~Asokan.
\newblock {Contextual Proximity Detection in the Face of Context-Manipulating
  Adversaries}.
\newblock Nov 2015.
\newblock arXiv report /1511.00905 http://arxiv.org/abs/1511.00905.

\bibitem{Project}
SourceForge.
\newblock {BlueProximity}.
\newblock https://sourceforge.net/projects/blueproxi-mity/.

\bibitem{Tefas2001}
A.~Tefas, C.~Kotropoulos, and L.~Pitas.
\newblock {Using Support Vector Machines to enhance the performance of elastic
  graph matching for frontal face authentication}.
\newblock {\em IEEE Transactions on Pattern Analysis and Machine Intelligence},
  23(7):735--746, 2001.

\bibitem{Tippenhauer2011}
N.~O. Tippenhauer, C.~P{\"{o}}pper, K.~B. Rasmussen, and S.~Capkun.
\newblock {On the requirements for successful GPS spoofing attacks}.
\newblock {\em ACM conference on Computer and communications security, CCS},
  page~75, 2011.

\bibitem{Truong2015}
H.~T.~T. Truong, X.~Gao, B.~Shrestha, N.~Saxena, N.~Asokan, and P.~Nurmi.
\newblock {Using contextual co-presence to strengthen Zero-Interaction
  Authentication: Design, integration and usability}.
\newblock {\em Pervasive and Mobile Computing}, 16:187--204, 2015.

\bibitem{Truong2014}
H.~T.~T. Truong, B.~Shrestha, N.~Saxena, N.~Asokan, and P.~Nurmi.
\newblock {Comparing and fusing different sensor modalities for relay attack
  resistance in Zero-Interaction Authentication}.
\newblock In {\em 2014 IEEE International Conference on Pervasive Computing and
  Communications (PerCom)}, pages 163--171. IEEE, mar 2014.

\bibitem{wagner2013device}
D.~T. Wagner, A.~Rice, and A.~R. Beresford.
\newblock Device analyzer: Understanding smartphone usage.
\newblock In {\em International Conference on Mobile and Ubiquitous Systems:
  Computing, Networking, and Services}, pages 195--208. Springer, 2013.

\bibitem{Watanabe2015}
T.~Watanabe, M.~Akiyama, and T.~Mori.
\newblock {RouteDetector: Sensor-based Positioning System That Exploits
  Spatio-Temporal Regularity of Human Mobility}.
\newblock 2015.

\bibitem{Xiao2016}
G.~Xiao, M.~Milanova, and M.~Xie.
\newblock {Secure behavioral biometric authentication with leap motion}.
\newblock In {\em 2016 4th International Symposium on Digital Forensic and
  Security (ISDFS)}, pages 112--118. IEEE, apr 2016.

\end{thebibliography}


\ifarxiv{
\appendix

\iftechnicalreport{

\section{Features}
\label{app:features}

Table~\ref{tab:features} shows the features extracted from segments of sensor readings. All features use the aggregated magnitude values of the three-dimensional vectors unless otherwise specified. Figure~\ref{fig:imp} shows the feature importances in the Random Forest algorithm. Some of the feature importances are 0.0, and this reflects successful feature selection. 

\begin{table*}[t]
  \centering
  \caption{Features used in movement classification.}

  \begin{tabular}{| l | l | c | l | }
    \hline
    $i$ & Feature name & Importance & Feature description \\
    \hline
    0 & AC\_time & 0.046 & Time lag with highest AC value, between [.1,2] seconds\\
    1 & AC\_val & 0.003 & Highest AC value\\
    2 & ACC\_var & 0.000 & Variance of 1s accelerometer magnitude\\
    3 & ACC\_p90 & 0.007 & 90th percentile of 1s accelerometer magnitude\\
    4 & ACC\_var\_long & 0.000 & Variance of 5s accelerometer magnitude\\
    5 & ACC\_p90\_long & 0.036 & 90th percentile of 5s accelerometer magnitude\\
    6 & GYRO\_var & 0.000 & Variance of 1s gyroscope magnitude\\
    7 & GYRO\_p90 & 0.002 & 90th percentile of gyroscope magnitude\\
    8 & GYRO\_var\_long & 0.000 & Variance of 5s gyroscope magnitude\\
    9 & GYRO\_p90\_long & 0.016 & 90th percentile of 5s gyroscope magnitude\\
    10 & GRAV\_std\_long & 0.001 & standard deviation of 5s gravity magnitude\\
    11 & ACC\_median & 0.004 & median of 1s accelerometer magnitude\\
    12 & ACC\_mean\_long & 0.019 & mean of 5s accelerometer magnitude\\
    13 & GYRO\_median & 0.001 & median of 1s gyroscope magnitude\\
    14 & GYRO\_mean\_long & 0.016 & mean of 5s gyroscope magnitude\\
    15 & GYRO\_p10 & 0.000 & 10th percentile of 1s gyroscope magnitude\\
    16 & GYRO\_p10\_long & 0.001 & 10th percentile of 5s gyroscope magnitude\\
    17 & ACC\_p10 & 0.001 & 10th percentile of accelerometer magnitude\\
    18 & ACC\_p10\_long & 0.014 & 10th percentile of 5s accelerometer magnitude\\
    19 & GYRO\_min & 0.001 & minimum of 1s gyroscope magnitude\\
    20 & GYRO\_max & 0.053 & maximum of 1s gyroscope magnitude\\
    21 & GYRO\_min\_long & 0.000 & minimum of 5s gyroscope magnitude\\
    22 & GYRO\_max\_long & 0.085 & maximum of 5s gyroscope magnitude\\
    23 & GYRO\_peak\_to\_peak & 0.234 & peak-to-peak difference for 1s gyroscope magnitude\\
    24 & GYRO\_peak\_to\_peak\_long & 0.035 & peak-to-peak difference for 5s gyroscope magnitude\\
    25 & GRAV\_min\_long & 0.000 & minimum of 5s gravity magnitude\\
    26 & GRAV\_max\_long & 0.000 & maximum of 5s gravity magnitude\\
    27 & GRAV\_peak\_to\_peak\_long & 0.000 & peak-to-peak difference for 5s gravity magnitude\\
    28 & GRAV\_p5\_long & 0.000 & 5th percentile of gravity magnitude\\
    29 & GRAV\_p95\_long & 0.000 & 95th percentile of gravity magnitude\\
    30 & GRAV\_interpercentile\_range\_long & 0.000 & interpercentile range(5-95) for 5s gravity magnitude\\
    31 & ACC\_std\_diff & 0.007 & std for 1s differenced accelerometer magnitudes\\
    32 & ACC\_std\_diff\_long & 0.147 & std for 5s differenced accelerometer magnitudes\\
    33 & ACC\_3\_std\_diff & 0.012 & std for 1s indiv. differenced and summed accelerometer values\\
    34 & ACC\_3\_std\_diff\_long & 0.257 & std for 5s indiv. differenced and summed accelerometer values\\
    \hline
  \end{tabular}
  \label{tab:features}
\end{table*}

\begin{figure}[tbh]

\centering
\includegraphics[width=.9\columnwidth]{pictures/feature_importances_rf_simple.png}
\caption{Random forest feature importances. The bars tell how often the features were selected to discriminate the classes and higher bars signify more important features. The bar heights sum to one.}
\label{fig:imp}
\end{figure}

\begin{table}[t]
  \centering

  \caption{Decision thresholds for Needleman-Wunsch algorithm. The r value for the OLS fit is shown in parentheses.}
  \label{table:dts}

  \begin{tabular}{| c | c | c |}
    \hline
    $\alpha$ & $D_L(L)$ & $D_{NW}(L)$  \\
    \hline

    $0.1$ & $1.571L - 0.412$ (r=$0.978$) & $22.310L+3.554$ (r=$1.000$) \\
    $0.2$ & $1.619L + 0.608$ (r=$0.980$) & $22.183L+1.424$ (r=$1.000$) \\
    $0.3$ & $1.784L + 0.601$ (r=$0.991$) & $22.194L+0.582$ (r=$1.000$) \\
    $0.4$ & $1.925L + 1.276$ (r=$0.991$) & $22.347L-0.584$ (r=$1.000$) \\
    $0.5$ & $2.004L + 1.386$ (r=$0.985$) & $21.552L+1.067$ (r=$1.000$) \\
    $0.6$ & $2.399L + 1.323$ (r=$0.992$) & $21.220L+0.785$ (r=$1.000$) \\
    $0.7$ & $2.496L + 1.364$ (r=$0.999$) & $21.407L-0.905$ (r=$0.999$) \\
    $0.8$ & $2.407L + 3.095$ (r=$0.991$) & $21.537L-2.388$ (r=$1.000$) \\
    $0.9$ & $2.806L + 3.418$ (r=$0.978$) & $21.187L-2.690$ (r=$1.000$) \\

    \hline
  \end{tabular}
  
\end{table}

Table~\ref{table:dts} shows the decision thresholds obtained by minimizing the error $\alpha \cdot FRR + (1-\alpha) \cdot FAR$, using Levenshtein distance and Needleman-Wunsch similarity. The slope remains largely constant between $0.1$ and $0.4$ and between $0.5$ and $0.9$ in NW. Both variables increase in the Levenshtein distance as $\alpha$ increases.

}
\else{

\begin{algorithm}                      
  \caption{Turn detection with angular data.}
  \label{alg1}                           
  \begin{algorithmic}                    
    \State{Heading angle $\alpha(t_0) = 0$ }
    \Repeat
    \State{Update heading angle $\alpha(t_i)$}
    \State Update \emph{rolling standard deviation} for last 2 seconds
    \State Update time when turn \emph{begins} ($t_{begin}$) and \emph{ends} ($t_{end}$)
    \If{turn \emph{ends}}
    \State $d = \alpha(t_{end}) - \alpha(t_{begin})$
    \State $n = |round(d / 15^{\circ})|$
    \If{$d > 0$}
    \State add $n$ symbols $R$
    \Else[$d < 0$]
    \State add $n$ symbols $L$
    \EndIf
    \EndIf
    \Until{app closed}
  \end{algorithmic}
\end{algorithm}

Algorithm~\ref{alg1} shows pseudocode for our turn detection algorithm. Gyroscope and accelerometer signals are re-sampled to 20Hz in \sysname. Axis-specific gyroscope noise is set to zero during stationary time ($std < 0.01$). Gyroscope events are realigned to the earth's frame by projecting along the gravity direction. The heading angle is acquired by integration (initial angle zero). Turns (yaw angle differences) are acquired continuously with Algorithm~\ref{alg1}.

\newpage

Trajectories are defined a sequences of $M$ (movement) and $L/R$ (left/right) primitives. Parameters used in \sysname are shown in Table~\ref{table:params}. Movement is classified with a logistic regression function every second. A hidden Markov model smooths the classified result into the sequence of most likely events. The sequence of movement and turn events are combined with turns taking precedence. Table~\ref{table:params} summarizes parameters used in \sysname. Trajectory comparison is done using the Needleman-Wunsch (NW) algorithm.

\begin{table}[h]
  \caption{Parameter used in \sysname.}
  \begin{center}
    \begin{tabular}{| l | c | }
      \hline
      Parameter & value   \\ \hline \hline
      Movement event time & $5$ seconds  \\ \hline
      $M$-event TPR & 98 \%  \\ \hline
      $S$-event TPR & 92 \%  \\ \hline \hline
      HMM prior $p(M \rightarrow M)$ & 99 \%  \\ \hline
      HMM prior $p(S \rightarrow S)$ & 99  \%  \\ \hline
      Movement classification frequency & $1$ second  \\ \hline \hline
      HMM Viterbi smoothing & at arrival to \ver  \\ \hline \hline
      Turn event granularity & $15^{\circ}$  \\ \hline  \hline
      Gyroscope flatten std thr. & $0.01^{\circ}$  \\ \hline
      Turn detection std thr. & $3^{\circ}$  \\ \hline
      Turn fine-tune std thr. & $1^{\circ}$  \\ \hline \hline
      Needleman-Wunsch match & $+1$  \\ \hline
      Needleman-Wu. mismatch & $-2$  \\ \hline
      Needleman-Wunsch gap & $-1$  \\ \hline
    \end{tabular}
  \end{center}
  \label{table:params}
  \end{table}

Table~\ref{table:dts} shows the \emph{initial} decision thresholds in NW obtained by minimizing the error rate $\alpha \cdot FRR + (1-\alpha) \cdot FAR$. A candidate path is accepted if its similarity to a reference path is higher than the decision threshold. The values are calculated for the \emph{pooled thresholds} (see Section~\ref{subsec:thresholds}).

\begin{table}[h]
  \centering

  \caption{Decision thresholds for Needleman-Wunsch algorithm, w.r.t. trajectory lengths in minutes. The r value for the OLS fit is shown in parentheses.}
  \label{table:dts}

  \begin{tabular}{| c | c |  }
    \hline
    $\alpha$ & $D(L)$  \\

    \hline


 $0.1$ & $10.171L+ 1.400$ ($r=0.998$) \\
 $0.2$ & $10.314L -0.600$ ($r=0.999$) \\
 $0.3$ & $10.143L -1.000$ ($r=0.999$) \\
 $0.4$ & $9.543L -0.067$ ($r=0.998$) \\
 $0.5$ & $9.686L -1.400$ ($r=0.998$) \\
 $0.6$ & $9.914L -2.533$ ($r=0.998$) \\
 $0.7$ & $9.771L -3.533$ ($r=0.996$) \\
 $0.8$ & $8.600L -1.600$ ($r=0.993$) \\
 $0.9$ & $7.914L -2.200$ ($r=0.989$) \\
    \hline
  \end{tabular}
  
\end{table}

\fi

\fi

\end{document}